\documentclass[useAMS, usenatbib]{mn2e}
\usepackage{epsfig}
\usepackage{amssymb}
\usepackage{amsmath}
\usepackage{graphicx}
\usepackage{tabularx}
\usepackage{rotating}
\usepackage{bm}
\usepackage{multirow}
\usepackage{verbatim} 
\usepackage{color}

\begin{document}

\title[WiggleZ 3-point correlation function]{The WiggleZ Dark Energy Survey: constraining galaxy bias
and cosmic growth with 3-point correlation functions}

\author[Mar\'in et al.]{\parbox[t]{\textwidth}{
Felipe A.\ Mar\'in$^{1}$\thanks{E-mail: fmarin@astro.swin.edu.au (FAM)}
    Chris Blake$^1$, 
    Gregory B.\ Poole$^{1,2}$,  
Cameron K. McBride$^{3}$,
      Sarah Brough$^4$, 
    Matthew Colless$^4$, 
    Carlos Contreras$^1$,
    Warrick Couch$^1$, 
    Darren J. Croton$^1$, 
    Scott Croom$^5$, 
    Tamara Davis$^6$, 
    Michael J.\ Drinkwater$^6$, 
    Karl Forster$^7$, 
    David Gilbank$^8$,  
    Mike Gladders$^9$, 
     Karl Glazebrook$^1$, 
    Ben Jelliffe$^5$, 
    Russell J.\ Jurek$^{10}$, 
    I-hui Li$^{11}$, 
    Barry Madore$^{12}$, 
    D.\ Christopher Martin$^7$, 
    Kevin Pimbblet$^{13}$, 
    Michael Pracy$^{1,5}$, 
    Rob Sharp$^{2,14}$,
    Emily Wisnioski$^{1,15}$, 
    David Woods$^{16}$, 
    Ted K.\ Wyder$^7$ and H.K.C. Yee$^{11}$} \\ \\
  $^1$ Centre for Astrophysics \& Supercomputing, Swinburne University of Technology, P.O. Box 218, Hawthorn, VIC 3122, Australia \\ 
    $^2$ School of Physics, University of Melbourne, Parksville, VIC 3010, Australia \\
  $^3$ Harvard-Smithsonian Center for Astrophysics, Cambridge, MA, USA\\
  $^4$ Australian Astronomical Observatory, P.O. Box 915, North Ryde, NSW 1670, Australia \\
  $^5$ Sydney Institute for Astronomy, School of Physics, University of Sydney, NSW 2006, Australia \\ 
  $^6$ School of Mathematics and Physics, University of Queensland, Brisbane, QLD 4072, Australia \\   
  $^7$ California Institute of Technology, MC 278-17, 1200 East California Boulevard, Pasadena, CA 91125, United States \\ 
  $^8$ South African Astronomical Observatory, PO Box 9 Observatory, 7935 South Africa \\
  $^9$ Department of Astronomy and Astrophysics, University of Chicago, 5640 South Ellis Avenue, Chicago, IL 60637, United States \\ 
  $^{10}$ Australia Telescope National Facility, CSIRO, Epping, NSW 1710, Australia \\ 
  $^{11}$ Department of Astronomy and Astrophysics, University of Toronto, 50 St.\ George Street, Toronto, ON M5S 3H4, Canada \\
  $^{12}$ Observatories of the Carnegie Institute of Washington, 813 Santa Barbara St., Pasadena, CA 91101, United States \\ 
  $^{13}$ School of Physics, Monash University, Clayton, VIC 3800, Australia \\ 
  $^{14}$ Research School of Astronomy \& Astrophysics, Australian National University, Weston Creek, ACT 2600, Australia \\ 
  $^{15}$ Max Planck Institut f\"{u}r extraterrestrische Physik, Giessenbachstra$\beta$e, D-85748 Garching, Germany\\
  $^{16}$ Department of Physics \& Astronomy, University of British Columbia, 6224 Agricultural Road, Vancouver, BC V6T 1Z1, Canada}

%\date{Accepted 1988 December 15. Received 1988 December 14; in original form 1988 October 11}
\date{Accepted for publication in MNRAS}

\maketitle

%\pagerange{\pageref{firstpage}--\pageref{lastpage}} \pubyear{2002}

%\label{firstpage}

\begin{abstract}
Higher-order statistics are a useful and complementary tool for measuring the clustering of galaxies,
containing information on the non-gaussian evolution and morphology of large-scale structure in the 
Universe. In this work we present measurements of 
the three-point correlation function (3PCF) for $187{,}000$ galaxies in the WiggleZ spectroscopic galaxy survey. 
We explore the WiggleZ 3PCF scale and shape dependence at three
different epochs $z=0.35$, 0.55 and 0.68, the highest redshifts where these measurements have been made to date.
Using N-body 
simulations  to predict the clustering of dark matter, we constrain the linear and non-linear 
bias parameters of WiggleZ galaxies with respect to dark matter, and marginalise over them to 
obtain constraints on  $\sigma_8(z)$, the variance
of perturbations on a scale of 8 $h^{-1}$Mpc and its evolution with redshift.
{\color{black} These measurements of $\sigma_8(z)$, which have 10-20\% accuracies, are consistent with the
predictions of the $\Lambda$CDM concordance cosmology and test this model in a new way.}
\end{abstract}


\section{Introduction}

In the current structure formation paradigm \citep[e.g.][]{press_shechter:74,white_rees:78,white_frenk:91,
berlind_weinberg:02}, 
galaxies form inside dark matter halos,
which evolved from small perturbations in the early universe. This allows us to connect the galaxy
field  to the overall matter distribution, and therefore to use large-scale galaxy  clustering 
to constrain  cosmological models and their  parameters (see for instance \citealt{peacock_etal:01, 
eisenstein_etal:05, percival_etal:07, kazin_etal:10, sanchez_etal:12} and references therein). 
This connection, 
however, is not a perfect one, since galaxy observables such as luminosity, colour, etc. are
also shaped by baryonic physics and environmental effects, with the consequence that different
types of galaxies have different clustering properties \citep{norberg_etal:01,zehavi_etal:05,zehavi_etal:11}, 
described as `galaxy bias'. 
 
The galaxy 2-point correlation functions have been the main tool to constrain cosmology using 
large-scale structure, because the shape of the 2-point clustering of matter depends on cosmological 
parameters such as the matter density, baryon fraction and neutrino mass.
In some cases it is possible to obtain these constraints marginalizing  over the bias of the 
galaxy  populations we use;
 an example is the cosmological constraints from Baryon Acoustic Oscillation (BAO) measurements
 \citep{eisenstein_etal:05,cole_etal:05,blake_etal:11bao, beutler_etal:11bao, sanchez_etal:12}
  where parameters such as the cosmic distance scale
   and the Hubble expansion rate $H(z)$ can be measured using only 
 the position of the BAO peak, which does not depend on the details of the galaxy population used, at first order.
 However, there are other cosmological parameters which cannot be constrained 
using  2-point galaxy clustering statistics only.

In particular, the amplitude of primordial perturbations, 
parameterized in the low-redshift universe by $\sigma_8(z)$, the r.m.s. of the matter density field in 8 $h^{-1}$Mpc spheres 
{\color{black} extrapolated to redshift $z$ by linear theory,} is degenerate
 with the details of the galaxy population, encoded in the large-scale linear galaxy bias parameter $b_1$.
On large scales these two parameters have the same effect on the  overall amplitude of the galaxy 2-point correlation functions; 
therefore,  one can only constrain the product $b_1\sigma_8$. 

Resolving this degeneracy requires use of another observable, or adoption of a particular
galaxy evolution model. For the first approach,  other observables such as lensing \citep[e.g.][]{fu_etal:08,lin_etal:11}, 
 or the mass function of galaxy clusters \citep[e.g.][]{eke_etal:96,rozo_etal:10,kilbinger_etal:12} have been used to constrain $\sigma_8$; but
 current constraints are  degenerate with other parameters such as the matter
 density $\Omega_{\rm{m}}$.
The second approach, selecting a particular galaxy 
population with a known evolution of its clustering,  allows  disentangling of the linear galaxy bias and $\sigma_8$, and 
has been studied in \cite{tojeiro_etal:12} using a passive-evolving luminous subsample of SDSS-I/II and SDSS-III survey
 galaxies. This method
 gives good constraints on $\sigma_8$; but  whereas it works well for their galaxy sample (of luminous red 
 galaxies), 
 for other galaxy surveys it might be difficult to find a suitable passive galaxy population. 

Another, complementary way to break these degeneracies is to measure three-point correlation functions (3PCF)
 of  the same galaxy dataset.  Two-point
statistics are only a complete description for Gaussian fields; but the late time large-scale structure, driven
by non-linear gravitational clustering, is strongly non-Gaussian \citep{bernardeau_etal:02}, and higher-order correlation 
functions thus encode additional information that can be  used to constrain galaxy population and cosmological 
models. The first measurements of the 3PCF were carried out in 
angular catalogues as a way to verify the hierarchical model of structure formation\citep{peebles_groth:75}, 
and more recently, in large-scale  
spectroscopic surveys such as 2dFGRS \citep{verde_etal:02, jing_borner:04, gaztanaga_etal:05},  and SDSS 
\citep{kayo_etal:04,nichol_etal:06,kulkarni_etal:07,gaztanaga_etal:08a,mcbride_etal:11b,marin:11,guo_etal:13}. 
The main goal of these measurements is to test theories of growth of structure and the predictions of
cosmological simulations, and to measure the biasing of the galaxies with respect to the dark
matter distribution.

In this work we present results of the measurement of the three-point
correlation function for a sample of 187,000  galaxies from the
 WiggleZ Dark Energy Survey \citep{drinkwater_etal:10}, which probes galaxies in the range $0.1<z<1.0$
 with a median redshift $z \sim 0.6$. 
 Using N-body simulations to study  dark matter statistics, we estimate the WiggleZ galaxy bias
 and thereby measure  $\sigma_8$. 
 These estimations have   been done in the past 
  \citep{gaztanaga_etal:05,ross_etal:06,marin:11,mcbride_etal:11b},
   but the fact that the WiggleZ survey spans such a large range of redshifts with a large overall
  volume allows us to split 
  our  galaxy sample in redshift slices and measure $\sigma_8$ as a function
 of redshift, hence constraining the large-scale structure growth history. 
 
 This paper is organized as follows: In $\S$2 we describe our survey and the simulations we use; in 
 $\S$3 we introduce the galaxy 3PCF and how it is measured, along with the model connecting
 galaxy and dark matter clustering. In $\S$4 we present the measurements of the WiggleZ 3PCF as
 a function of scale and shapes. In $\S$5 we discuss constraints on the galaxy bias and $\sigma_8$ as 
 a function of redshift. In $\S6$ we summarize our findings. 
 {\color{black} We note that  a fiducial flat $\Lambda$CDM 
 cosmological model 
 with matter density $\Omega_{\rm{m}} = 0.27$ and Hubble parameter $H_0=$ 100 $h$ km s$^{-1}$ Mpc$^{-1}$ 
 with $h=0.7$
 is used
 throughout this paper to convert redshifts to distances, which are measured in $h^{-1}$Mpc.}

\section{Data and Simulations}
\label{sec:data}

\begin{figure*}
\includegraphics[width=17cm]{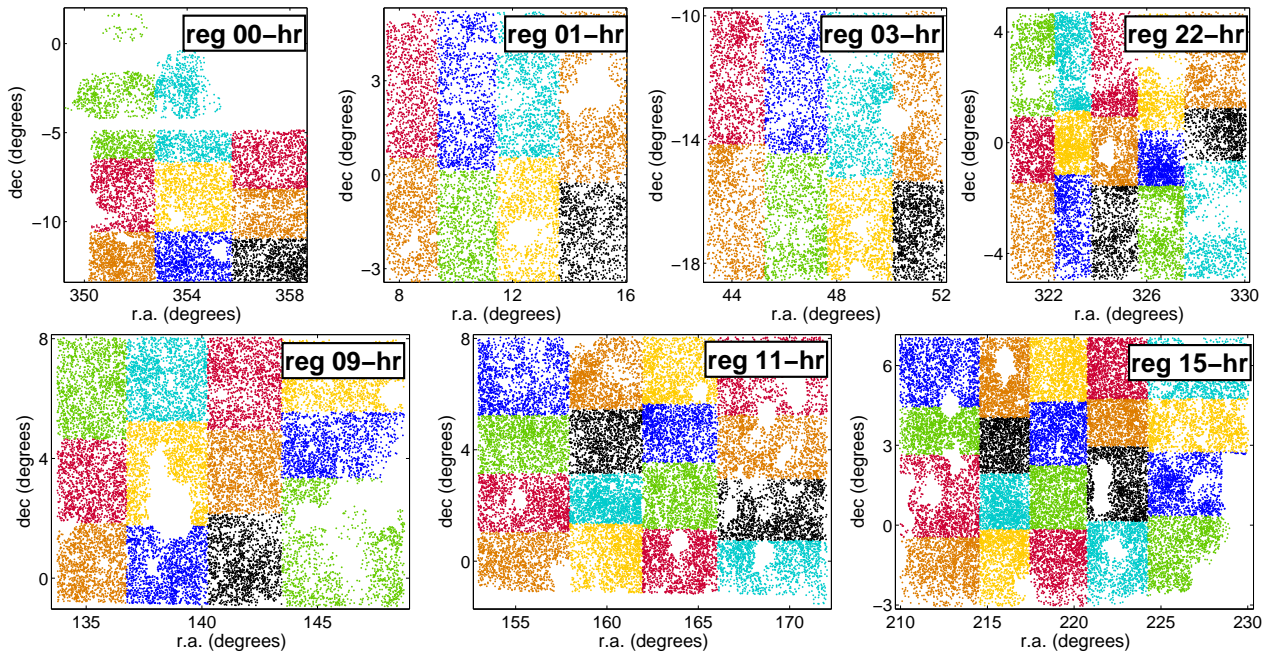}
\caption{\label{fig:angsf}
Angular distribution of WiggleZ galaxies. The top four regions correspond to those
WiggleZ galaxies in the RCS2 footprint; the bottom three regions to the ones obtained using SDSS. 
Colours correspond to sub-regions containing the same effective area.}
\end{figure*}

\subsection{The WiggleZ Galaxy Survey}
\label{sec:wzdesc}

The WiggleZ Dark Energy Survey \citep{drinkwater_etal:10} is a large-scale galaxy redshift survey performed over 
276 nights with the AAOmega spectrograph \citep{sharp_etal:06} on the 3.9m Anglo-Australian Telescope. 
With a area coverage
of 816 deg$^2$, this survey has mapped $240,000$ bright emission-line galaxies over a cosmic volume of 
approximately 1 Gpc$^3$.  

Target galaxies in
seven different regions  were chosen using
UV photometric data from the \emph{GALEX} survey \citep{martin_etal:05}
 matched with optical photometry from the
Sloan Digital Sky Survey (SDSS DR4,  \citealt{adelman_etal:06}) for regions in the Northern Galactic
Cap (9-hr, 11-hr and 15-hr), and 
from the Red-Sequence Cluster Survey 2 (RCS2, \citealt{gilbank_etal:11}) for those regions in the 
Southern Galactic Cap (0-hr, 1-hr, 3-hr, 22-hr). 
The selection criteria consisted of applying magnitude and colour cuts \citep{drinkwater_etal:10} in order to select
 star-forming galaxies with bright emission lines with a redshift distribution centered around $z\sim0.6$.
 The selected galaxies were observed in 1-hour exposures using the 
 AAOmega spectrograph, and their redshifts were estimated from strong emission lines.  
 
To study the evolution of
the bias and $\sigma_8$ with cosmological time,  we use three overlapping redshift slices
$[0.1,0.5]$, $[0.4,0.8]$ and $[0.6,1.0]$. We estimate the effective redshift of each sample by averaging the redshifts
of galaxy pairs at the distances covered by our study, i.e. from 10 to 100 $h^{-1}$Mpc; we find that the
effective redshifts for the closest, middle and farthest slices are $z_{\rm{eff}}$=[0.35, 0.55, 0.68] respectively.
Table \ref{tab:samples} shows details of the samples used. 

 \begin{table}
\begin{minipage}{80mm}
\centering
 \caption{\label{tab:samples}Number of galaxies in  WiggleZ regions used in this paper.}
  \begin{tabular}{l c c c c}
  \hline
    redshift& $[0.1,0.5]$ & $[0.4,0.8]$ &  $[0.6,1.0]$ &JK subregions\\
    $z_{\rm{eff}}$ &$0.35$&$0.55$&$0.68$&$N_i^{JK}$\\
 \hline
 00-hr & 6601 & 10698 & 8774&9\\
 01-hr & 6038 & 9437 & 7880&8\\
 03-hr & 6492 & 10241 & 8756&8\\
  22-hr & 13508 & 16146 & 11024&15\\
 09-hr & 10106 & 18978 & 11424&12\\
 11-hr & 13603 & 23940 & 13919&16\\
 15-hr & 14517 & 30015 & 19471&20\\
 all regions & 74440 & 119455 & 81248&88\\
 \hline
\end{tabular}
\end{minipage}
\end{table}

 Figure \ref{fig:angsf}  shows the angular distribution of galaxies in the regions considered. 
We show the targets in r.a.-dec. coordinates, where it can be seen that the angular completeness varies considerably 
between regions,  due to  masking of {\color{black}bright stars}, the availability of input GALEX imaging, 
and differences in the accumulated observation time within each  region and between regions.
If not taken into account properly by modelling the angular selection function, these non-uniformities
may lead to artificial structures, different from what we can expect from cosmic variance.
Several studies, such as \cite{gaztanaga_etal:05, nichol_etal:06, mcbride_etal:11a,norberg_etal:11}, agree that 
 higher-order correlation functions are more sensitive to these effects  than the 2-point function. 
However, as we 
 describe below in $\S$3.2, we conclude that our modelling of the angular completeness is adequate to carry out
 our analyses.
 
\begin{figure}
\includegraphics[width=18pc]{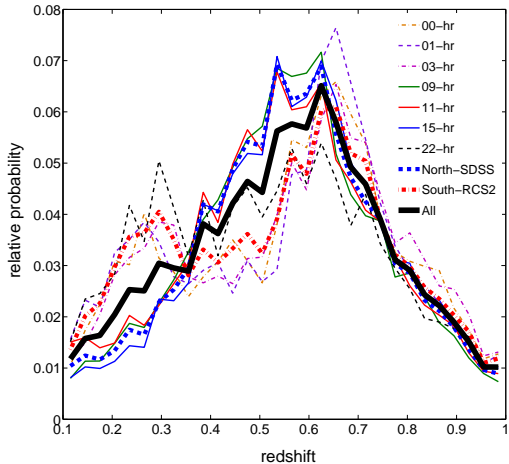}
\caption{\label{fig:radsf}
Redshift distribution of WiggleZ galaxies. Thin lines correspond to the radial selection function for each individual angular region. 
Thick lines represent 
the redshift distribution of all WiggleZ galaxies (black), 
WiggleZ galaxies in the North Cap (blue) and WiggleZ galaxies in the South Cap (red).}
\end{figure}

Figure \ref{fig:radsf} shows the redshift distribution of the
 different regions, peaking at $z\sim0.6$, but extending  to redshift $z\sim1.0$.
 The variable number density with redshift determines the effective redshifts of our samples
 as measured above.  
It can also be seen  that the average distribution of galaxies varies  between regions. 
  This is partly explained by cosmic variance, but also the selection functions of SDSS and RCS2 galaxies differ considerably at low redshifts, 
  owing to the available colours for galaxy selection from the input catalogues  \citep{drinkwater_etal:10}.
   To deal with these issues we model  the {\color{black} angular coverage and redshift distributions}
  in each survey region individually \citep{blake_etal:10sf}.

 \subsection{The GiggleZ simulations}
 \label{sec:gigglez}
 
In order to measure galaxy bias we need to model the underlying dark 
matter correlations. For the 2-point functions there exists a large literature of models \citep[e.g.][and references therein]{peebles:80, 
kaiser:87, cooray_sheth:02,bernardeau_etal:02,smith_etal:03}, 
but in the case of the higher-order correlations,   
modelling has focused mostly on the  large-scale behaviour \citep[e.g][]{jing_borner:97,bernardeau_etal:02}, although
there have been efforts to model the non-linear, small scales \citep[e.g][]{yang_etal:02, fosalba_etal:05}.
 Most importantly, for the 3PCF there is no 
satisfactory treatment of redshift-space distortions, although some attempts have been made for the bispectrum, the 
Fourier  transform of the 3PCF, by  \cite{scoccimarro_etal:99, smith_etal:08}, which are valid for limited range of scales.
Therefore, 
 as  has been done in previous works \cite{gaztanaga_etal:05, marin:11,mcbride_etal:11b},
 we will obtain constraints on the galaxy bias and cosmological parameters by comparing the WiggleZ galaxy 3PCF with the 
dark matter correlations measured in  N-body simulations, {\color{black} which include the full set of non-linear effects.}

We measured the dark matter correlation functions using  the Gigaparsec
WiggleZ Survey simulations   \citep[GiggleZ,][]{poole_etal:12}, 
which have been generated in support of  WiggleZ science.
 In a 1 $h^{-1}$Gpc$^3$ periodic cube, $2160^3$
dark matter particles with individual masses of $m_p=7.5\times10^9$ $h^{-1}$$M_\odot$ were evolved
using a flat $\Lambda$CDM model, with cosmological parameters from WMAP5 results \citep{komatsu_etal:09}. 
In order to compare the correlation functions of WiggleZ galaxies and dark matter, we measured them in
snapshots of the simulation at the same effective redshifts as  the WiggleZ subsamples. 

Dark matter halos are identified in two steps \citep{springel_etal:01b}: Firstly, using a 
 friends-of-friends algorithm
 with a linking length of $l=0.2 $ {\color{black} times the mean particle separation}, bound structures are found (parent halos). 
 Secondly, given the 
 high resolution of our simulation, we were able to find gravitationally bound substructures inside these parent halos.
  From the main
 halo catalog, we create {\color{black}subsets} ordered by   maximum circular velocity (which we use as a proxy for halo mass) with
 the same mean number density as the WiggleZ galaxies $n\sim 2.5\times 10^{-4}$ ($h^{-1}$Mpc)$^{-3}$.
These  halo catalogues are used 
 to carry out consistency checks in our phenomenological model to estimate the bias and cosmological parameters. 

 \section{The galaxy  three-point correlation function}
 \label{sec:3pcf}

\begin{figure}
\includegraphics[width=22pc]{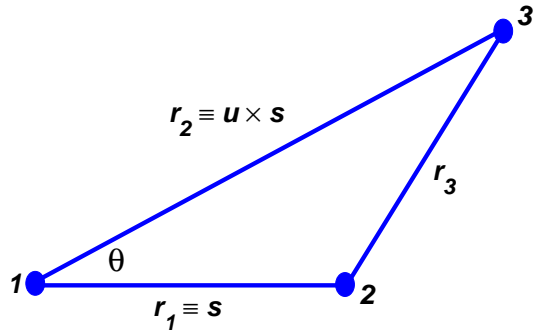}
\caption{\label{fig:tri}
Parameterization of triangles for calculation of correlation functions.}
\end{figure}

 \subsection{Definitions and Methods}
 
 The galaxy $n$-point correlation functions
are the average of correlated galaxy overdensity  $\delta_{\rm{gal}}$ measured
at $n$ different points \citep{peebles:80}. Whereas the two-point correlation function $\xi(r)$ (2PCF) 
allows us to estimate the probability of finding pairs with a {\color{black} separation} $r$, 
 the three-point correlation function $\zeta(r_{1},r_{2},r_{3})$ (3PCF)
 describes the probabilities of finding triplets with galaxies as vertices. 
 The joint probability of finding three objects in three infinitesimal volumes
$dV_1$, $dV_2$, and $dV_3$ is given by the {\color{black} `full' three-point correlation
function:} \citep{peebles_groth:75,peebles:80}
\begin{eqnarray}\label{eq:P3}
P &= &[1+\xi(r_{1})+\xi(r_{2})+\xi(r_{3})+\zeta(r_{1},r_{2},r_{3})]~\nonumber\\
&~ &\times \bar{n}^3 dV_1 dV_2 dV_3 ~,
\end{eqnarray}
where $\bar{n}$ is the mean density of objects, $\xi$ is the
2PCF, and $\zeta$ is the {\color{black} reduced or  `connected' 3PCF . In other words, this means that
the probability to find galaxies in  a particular triangular configuration has contributions from triplets found by 
random chance, plus contributions from correlated pairs plus the third point found at random (the $\xi$ terms), and lastly by
intrinsically correlated triplets (the $\zeta$ term). In the dark matter or galaxy field the 2PCF and 3PCF are given by:}
\begin{eqnarray}
\xi(r) &=& \langle\delta(\mathbf{x_1})\delta(\mathbf{x+r})\rangle \\
\zeta(r_{1},r_{2},r_{3})&=&\langle\delta(\mathbf{x_1})\delta(\mathbf{x_2})\delta(\mathbf{x_3})\rangle,
\end{eqnarray}
where $\delta$ is the fractional overdensity of objects (galaxies, halos or dark matter particles)
or the continuous field studied, and $\mathbf{x_1}$, $\mathbf{x_2}$ and  $\mathbf{x_3}$ form a closed triangle (see Figure \ref{fig:tri}).
The triangle sides $r_i$  
are the distances between objects in the triplet; thus the 3PCF depends
upon the scales and
shapes of spatial structures   
\citep{barriga_gaztanaga:02, 
sefusatti_scoccimarro:05,gaztanaga_scoccimarro:05,marin_etal:08}.
Since the ratio $\zeta/\xi^2$  is both predicted on large scales from perturbation theory \citep{bernardeau_etal:02} 
and measured  to be close to
unity over a large range of length scales, even though $\xi$ and $\zeta$ each    
vary by orders of magnitude \citep{peebles:80}, we will often present results using the `reduced' (or normalized)
3PCF, 
\begin{equation}
Q(s,u,\theta) \equiv \nonumber \frac{\zeta(s,u,\theta)}{\xi(r_{1})\xi(r_{2})+ \xi(r_{2})\xi(r_{3})+\xi(r_{3})\xi(r_{1})}~.
\end{equation}
Here, $s\equiv r_{1}$ sets the scale size of the triangle, and the shape
parameters are given by the ratio of two sides of the triangle, 
$u \equiv r_{2}/r_{1}$, and the angle between those two sides,
$\theta=\cos^{-1}(\hat{r}_{1}\cdot \hat{r}_{2})$, 
where $\hat{r}_{1}$, $\hat{r}_{2}$ are unit vectors in the directions of 
those sides. 
The reduced 3PCF is also better suited for visualizing the growth of non-gaussian
structure and the shape dependence of clustering  than $\zeta$.
On the other hand on large scales, since $\xi \rightarrow 0$, 
the ratio $\zeta/\xi^2$  becomes very unstable and its errors non-Gaussian, with
the consequences of  overestimating the
covariances, diminishing the overall signal-to-noise ratio and  introducing a systematic deviation
in the confidence intervals of the fitted parameters, {\color{black} therefore we use $\zeta$ and not $Q$ for model
fits.}

In this work we measure correlation functions for triangles with base side $s=10$, 15, 20 and 30 $h^{-1}$Mpc, with
the shape parameters $u=1.0$, 2.0, 3.0 and 10 equally spaced bins in $\theta$. In total  we measure correlations
for 120 triangular configurations. 

\subsection{Measuring Correlation Functions}

\begin{figure}
\includegraphics[width=19pc]{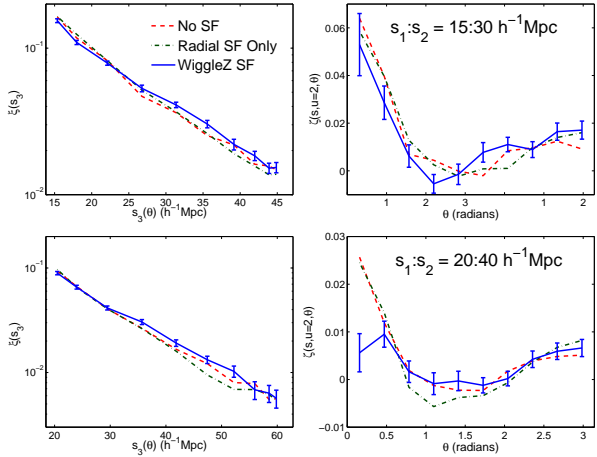}
\caption{\label{fig:sfcomp}
Correlation function measurements for a GiggleZ halo catalog. Selecting a set of triangular configurations, with $s=15$ $h^{-1}$Mpc  (top)
and $s=20$ $h^{-1}$Mpc  (bottom) and $u=2$, we plot 
the redshift-space reduced 2PCF of the third side  $\xi(s_3(\theta))$ (left panels) and connected 3PCF, $\zeta$ (right panels) 
of a selected DM halo 
catalogues from the GiggleZ simulation with similar clustering as WiggleZ galaxies, showing the effects of radial and angular selection functions.}
\end{figure}

We measure first the 2PCF and 3PCF in each WiggleZ region (i.e in angular and redshift cuts). 
For a particular WiggleZ region, we calculate
the 2PCF using the estimator of \cite{landy_szalay:92}, 
\begin{equation}
\xi = \frac{DD-2DR+RR}{RR}.
\end{equation}
Here, $DD$ is the number of data pairs normalized by $N_D\times
N_{D}/2$, $DR$ is the number of pairs using data and random catalogues
normalized by $N_DN_R$, and $RR$ is the number of 
random pairs normalized by
$N_R\times N_R/2$, where $N_D$ and $N_R$ are the number of points in the data and
in the random catalog of the region, respectively.  The 3PCF is calculated using the
\cite{szapudi_szalay:98} estimator:
\begin{equation}
\zeta=\frac{DDD-3DDR+3DRR-RRR}{RRR},
\end{equation}
where $DDD$, the number of data triplets, is normalized by $N_D^3/6$,
and $RRR$, the random data triplets, is normalized by $N_R^3/6$. $DDR$ is
normalized by $N_D^2N_R/2$, and $DRR$ by $N_DN_R^2/2$. Due to the low density
of our galaxies $n\sim 2.5\times 10^{-4} \, h^3$ Mpc$^{-3}$, we are limited by shot noise on these large
scales and consequently the application of FKP weighting \citep{feldman_etal:94} to the pair and triplet counts does not affect the results.
To estimate the number of pairs and triplets, we use the {\it ntropy-npoint} software, an
exact n-point calculator  which 
uses a kd-tree framework with true parallel capability and 
enhanced routine performance \citep{gardner_etal:07,mcbride_etal:11a}.

The random catalogues were built using the methods described by \cite{blake_etal:10sf},
 which estimate the angular and radial selection function of each survey region   
  due to survey geometry and incompleteness in the parent photometry and 
 spectroscopic follow-up. This modelling process produces a series of Monte Carlo random realisations of
 the angular and redshift catalogue in each region, which is used in our  correlation function estimations. 
In this paper, we measure the 3PCF using 10 random catalogues for each region, 
with $N_R=4N_D$ for each of the random catalogues for the 
intermediate scales ($s=10,15$ $h^{-1}$Mpc), and $N_R=N_D$ for the largest scales
 ($s=20,30$ $h^{-1}$Mpc). 
 For our choice of binning (resolution) of the triangles, we use the same scheme as 
\cite{marin:11, mcbride_etal:11a}: first, we select the central $s$, $u$ and $\theta$, 
and their corresponding side lengths in redshift space $s_i$, with $i=1,2,3$. Then, to calculate the 2PCF and 3PCF,
 we accept triangles with sides between  $(1 - 0.075)s_i< r_i < (1 + 0.075)s_i$, implying a 15\% binning resolution. 
 This is a higher resolution than used for the LRGs 
\citep{marin:11} but is justified by the higher number density of the WiggleZ galaxies.

In Figure \ref{fig:sfcomp} we explore the effects of the radial and angular selection functions on the 2PCF and
connected 3PCF  $\zeta$ for  a selection of configurations ($s=15,20$ $h^{-1}$Mpc, $u=2$).
Using a dark matter halo catalog from the GiggleZ simulation 
 that has a similar 2-point clustering as our WiggleZ sample at $z=0.55$,
we create three different kinds of mock catalogues with the same geometry as the survey: the first group (in red, 
dashed line) does not
 include any radial or angular selection effects apart from the large-scale boundaries
of the WiggleZ regions. 
 Green (dotted-dashed) lines are measurements from mocks with the same radial selection function
of WiggleZ galaxies at $z_{\rm eff} = 0.55$. Blue (solid) lines  are mocks with the same angular 
and radial selection function of WiggleZ galaxies. 
In general, the measurements of the correlation functions
using the different mocks are very similar, signaling that our selection functions and random catalogues allow us
to recover the intrinsic correlations of the galaxy field.

\subsection{Galaxy Bias Model}

Since different types of galaxies form inside different dark matter halos, 
they are an imperfect tracer of the overall dark matter distribution \citep{bardeen_etal:86,
cooray_sheth:02, berlind_weinberg:02}, 
and their $n$-point correlations will differ as well. 
 Many  models of this galaxy bias have been proposed, and an accepted working model on large scales is the
deterministic and local bias formalism, where we relate real space galaxy overdensity $\delta_{\rm{gal}}$ 
to the underlying matter density $\delta_{\rm{m}}$ \citep{fry_gaztanaga:93,frieman_gaztanaga:94}:
\begin{equation}
\delta_{\rm{gal}} = b_1\delta_{\rm{m}}+\frac{b_2}{2}\delta_{\rm{m}}^2+...\\
\label{delta-bias}
\end{equation}
up to second order, where $\delta_{\rm{gal}}$ and $\delta_{\rm{m}}$ are the local galaxy and 
 matter overdensities smoothed over some scale $R$. 
To leading order, this bias prescription leads to a relation 
between the galaxy and matter  2PCF and connected 3PCF.
Following \cite{pan_szapudi:05}, 
leading order  Perturbation Theory \citep{bernardeau_etal:02} shows that
if we fix all cosmological parameters except the overall amplitude of the initial spectrum 
of perturbations  characterized by $\sigma_8$, then there is a degeneracy between the effect of
this parameter and the bias on the 2-point and 
3-point functions. The relations between matter and galaxy correlations in this model are:
\begin{eqnarray}
\xi_{\rm{gal}}(r) &=&(\sigma_8/\sigma_{\rm{8,fid}})^2 b_1^2\xi_{\rm{m}}(r) \\
\zeta_{\rm{gal}}(r_{12},r_{23},r_{31}) &=&(\sigma_8/\sigma_{\rm{8,fid}})^4[ b_1^3\zeta_{\rm{m}}(r_{12},r_{23},r_{31}) + \nonumber \\
&& b_2b_1^2(\xi_{\rm{m}}(r_{12})\xi_{\rm{m}}(r_{23}) + {\rm{perm.}})].
\label{eq:realbias}
\end{eqnarray}
In observations we measure the  correlation functions in redshift space, where the `real' correlations are distorted
by peculiar velocities (with respect to the Hubble expansion), which on large scales depend on the growth rate of 
perturbations $f\approx\Omega_{\rm{m}}(z)^{0.55}$ and on galaxy  bias. For the 2-point function we have that
$\xi_{\rm{z-space}} \sim f_2 \xi_{\rm{r-space}}$, with \citep{kaiser:87}:
\begin{equation}
f_2 = 1 + \frac{2}{3}\left(\frac{f}{b_1}\right) +  \frac{1}{5}\left(\frac{f}{b_1}\right)^2.
\end{equation}
In the case of the 3PCF, there is also an effect from redshift space
distortions of similar order \citep{scoccimarro_etal:99}. However, it
depends not only on the linear bias and $f$, but also on the non-linear bias and the shape and
scale parameters of the triangle observed. As mentioned before, 
analytical models of these distortions have been 
proposed for the bispectrum on large scales \citep{scoccimarro_etal:99} where its validity is limited. 
On small, non-linear scales \citep{smith_etal:08} the transformation back to configuration space
is challenging (six-dimensional integrals) and numerically intractable. 

For these reasons we opt to use an empirical model that has been used by \cite{pan_szapudi:05, gaztanaga_etal:05}
{\color{black} in analyses of the 2dFGRS galaxies.} 
We  use the $N$-body measurements 
of the correlation functions in redshift space in equations (7) and (8), 
replacing $\xi_{\rm{r-space}}\rightarrow\xi_{\rm{z-space}}$
and $\zeta_{\rm{r-space}}\rightarrow\zeta_{\rm{z-space}}$ at the different effective redshifts of our WiggleZ slices.
 Given the low signal-to-noise ratio of the 3PCF measurements, this is justified by the fact that
for low bias tracers {\color{black}(such as the 2dFGRS galaxies studied in the papers mentioned above)}
 with $b\sim1$, $f_{2,b_1}\approx f_{2,b_1=1}$, with differences of the order of 10\% when
$(b_1-1)=0.3$, smaller than the statistical error in our measurements {\color{black} (of the order 20\%-30\%)}; 
the impact on the constraints on $\sigma_8$  is slightly lower. 
{\color{black} Since we are using N-body simulation measurements to compare to our galaxy correlations, the only cosmological
parameter we modify is $\sigma_8$ through the ratio $\sigma_8/\sigma_{\rm{8,fid}}$ in eqs. (7) and (8).}

 \begin{figure}
\includegraphics[width=8cm]{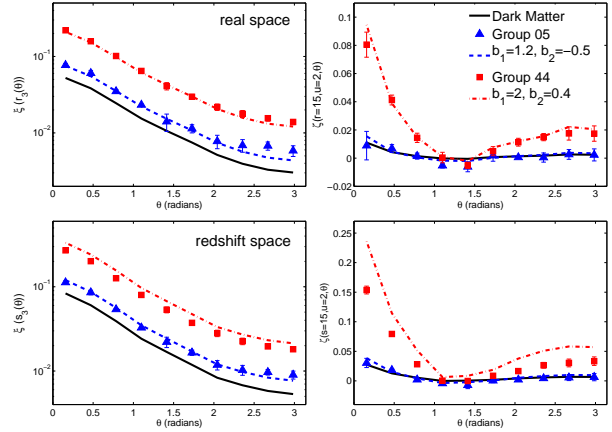}
\caption{\label{fig:gh0p6}
The 2PCF $\xi(s_3(\theta))$  (left) and connected 3PCF $\zeta$ (right) of two samples (groups) of halos
 from the GiggleZ simulations in real space (top) and
redshift space (bottom) at a snapshot  $z=0.6$.
 Black lines are results from the dark matter particles from the simulation, blue triangles 
display measurements for a halo {\color{black}sample (`group 05')} with average mass $4.8\times10^{11}$ $h^{-1} M_\odot$; 
red squares show results for a halo {\color{black}sample (`group 44')} with average mass   $6.9\times10^{12}$ $h^{-1}M_\odot$. Blue (short 
dashed) and red (dot dashed) are biased DM for low mass and high mass halos respectively (bias parameters are
listed in the figures). }
\end{figure}

The justification of this model is  illustrated   by Figure \ref{fig:gh0p6},  
where we take two halo {\color{black}samples} (groups) from the GiggleZ 
simulation at redshift $z=0.6$, with a similar number density as WiggleZ galaxies $n_h=2.5\times10^{-4} $ 
($h^{-1}$Mpc)$^{-3}$. The first {\color{black}sample, `group 05',} is composed   
of low mass halos, with clustering similar to that of WiggleZ galaxies, 
and the second sample,  {\color{black}`group 44',} consists of
very massive halos.  We measured their  correlation functions in real  space and then estimated their linear
and non-linear bias
 parameters {\color{black} by comparison with the matter correlation functions measured in the GiggleZ simulation, using eqs. (7) and (8) for 
 a fixed $\sigma_8(z=0)=0.812$ (see $\S$5.1)}. In the case of the low bias sample, 
 $\chi^2_{\rm r,g05}/$dof$=0.87$, and for the high mass sample  $\chi^2_{\rm r,g44}/$dof$=0.96$
  We then performed measurements 
 in redshift space; we observe that in the case of the low-bias tracers, 
using the same bias in redshift space fits well their redshift space  2PCF  and 3PCF, with
$\chi^2_{\rm z,g05}/$dof$=1.07$, but 
 the same can not be said for the high-mass halos where {\color{black} $\chi^2_{\rm z,g44}/$dof$=22.44$}.
  Since WiggleZ galaxies have a low linear bias 
 $b_1 \sim 1$ \citep{blake_etal:11f, contreras_etal:12w}, the approach we take
 is adequate for obtaining measurements of the linear and nonlinear bias parameters and $\sigma_8$ as a function of
 redshift.

\section{Results}

In this section we present the measurements of the 2PCF $\xi$, connected 3PCF $\zeta(s,u,\theta)$ and 
reduced  3PCF, $Q(s,u,\theta)$, of WiggleZ galaxies
for a range of scales and shapes at different redshifts. We explore differences between regions, evolution
with redshift, and our estimation of statistical errors and covariance.

\subsection{Building the WiggleZ Survey correlations}

\begin{figure}
\begin{minipage}{18pc}
\includegraphics[width=20pc]{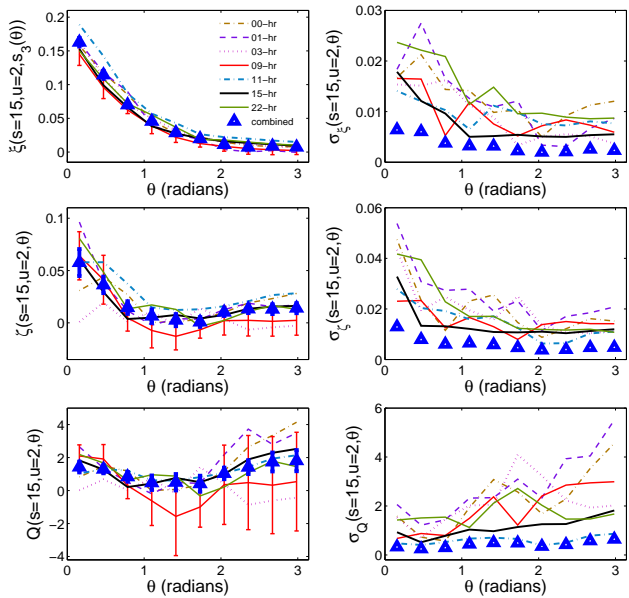}
\caption{\label{fig:allregss15u2}
The 2PCF $\xi(s_3)$ (top left), the connected 3PCF $\zeta$ (middle left) and the reduced 3PCF $Q$ (bottom left) 
of each WiggleZ region in the $z_{\rm{eff}}=0.55$ slice, for 
triangles with $s_1=15$ $h^{-1}$Mpc and $u=2$.  In the right-hand plots we show the corresponding diagonal errors.}
\end{minipage}\hspace{1.5pc}%
\end{figure}

In Figure \ref{fig:allregss15u2} we show measurements of the correlation functions in the $z_{\rm{eff}}=0.55$ sample
for each WiggleZ region.
Measurements in different regions are consistent within the statistical errors from cosmic variance and shot noise. 
On these scales we notice how small differences in $\xi$ and $\zeta$ translate
to larger discrepancies in $Q$. 
 The noisiest 3-point functions are obtained in the smallest
regions (in terms of volume), in this case the regions overlapping with the RCS2 survey.
We will build a `combined' set of correlation functions calculated by optimally weighting individual 
contributions of the regions.

\begin{figure*}
\includegraphics[width=15cm]{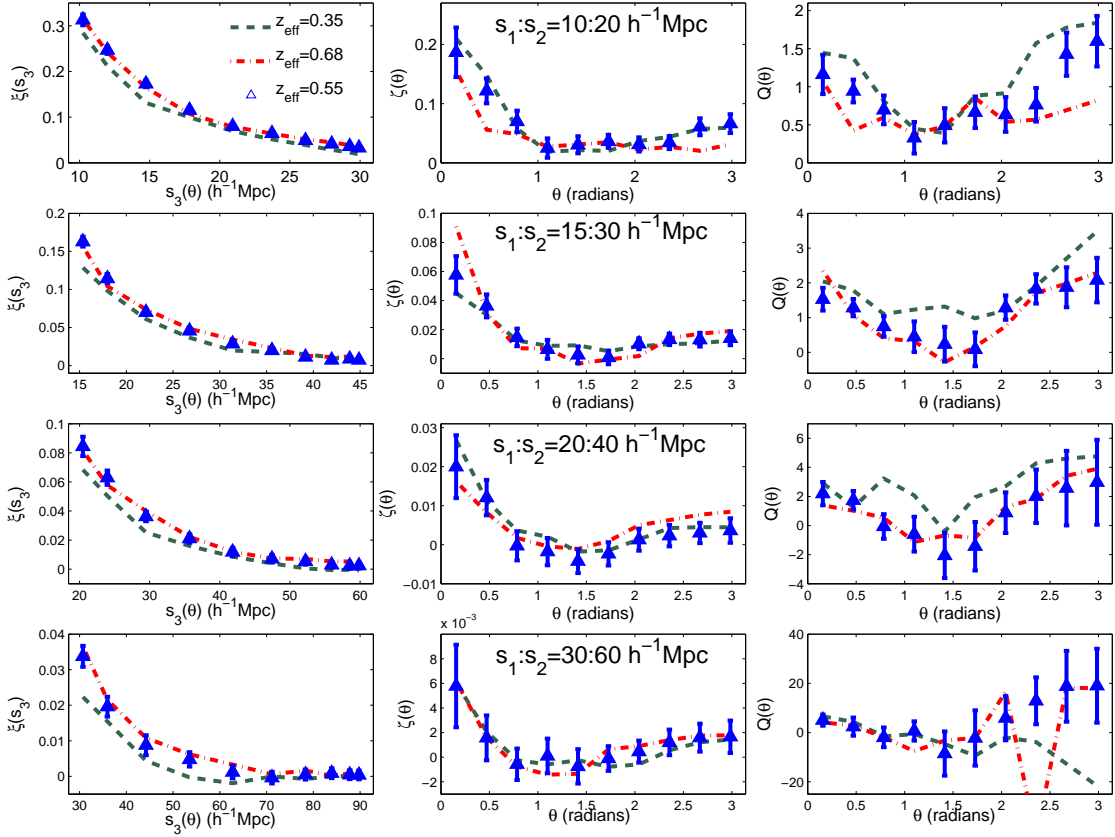}
\caption{\label{fig:wz3pt0p6}
The combined redshift-space 2PCF $\xi(s_3(\theta))$ (left), connected 3PCF $\zeta(s,u=2,\theta)$ (middle), and 
reduced 3PCF $Q(s,u=2,\theta)$ (right) of WiggleZ galaxies in the $z_{\rm{eff}}=0.55$ slice (blue triangles and error bars),  
in the slice at $z_{\rm{eff}}=0.35$ (green dashed line) and in the $z_{\rm{eff}}=0.68$ slice (red dashed-dotted line).
{\color{black} Different rows cover the range of scales of the triangles.}
 Errors have been determined by jack-knife resampling.}
\end{figure*}

To measure the diagonal errors and covariance matrices, ideally we require a large number
of  mock galaxy catalogues whose correlation functions have similar amplitude and shape dependence to
the one observed in our data. However,
this is not currently tractable for WiggleZ galaxies due to their low bias, which would necessitate many 
high-resolution simulations in a cosmological volume.
 In past studies we have used lognormal realisations \citep{blake_etal:10sf}
 to generate a large number of mocks suited to match the WiggleZ 2-point amplitudes. Unfortunately, by their
 construction (from generating points with a particular 2-point distribution),
  these are not capable of reproducing the higher-order clustering of galaxies. 
  
 In this work, error measurements in each region are calculated from jack-knife resampling
  \citep{zehavi_etal:05, mcbride_etal:11a,norberg_etal:11}. {\color{black} In this method 
  we divide the whole volume of the sample in identical subsamples $i=1...$ $N$, and we then measure correlation functions
  for the whole volume minus the $i$-th subsample $N$ times to get a set of $N$ correlated measurements.} 
In our case, 
in all WiggleZ regions we take
 subregions
  of equal area (weighted by sky  completeness), with an equivalent physical size of approx.
 $120\times150\times900$ ($h^{-1}$Mpc)$^{3}$ at $z=0.55$.
  The angular cut has the same area independent
 of region,  as can be seen in Figure  
 \ref{fig:angsf},  thus  some WiggleZ regions have more JK subregions than others, 
 depending on their total area coverage. 
 
In order to obtain the  JK variance in each region, we measure each $X_i$ statistic
(where $X_i$ can be the 2PCF or the 3PCF), subtracting
one of the JK subregions in turn. Then  we calculate the variance $\sigma_{X_i}$ of the individual WiggleZ region as 
\begin{equation}
\sigma_{X_i}^2 = \frac{N^{JK}_i-1}{N_i^{JK}}\sum_{j=1}^{N_i^{JK}}(X_{i,j}^2 - \langle X_i\rangle^2);
\end{equation}
where $N^{JK}_i$ is the number of jack-knife subsamples,  in region $i$ (see Table 1). Then,
we calculate the correlations in the overall survey using inverse-variance weighting. For 
the statistic $X_{\rm{comb}}$, this is calculated as 
\begin{equation}\label{eq:xcomb}
  X_{\rm{comb}}=  \left(\sum_{i=1}^{N_{reg}}  \frac{1}{\sigma_{X_i}^2}\right)^{-1}   \left(\sum_{i=1}^{N_{reg}} \frac{X_{i}}{\sigma_{X_i}^2}\right) 
\end{equation}
where  $\sigma_{X_i}^2$ is the variance of the statistic in the WiggleZ subregion (calculated in equation 10), 
taken from the jack-knife resampling method.
 $N_{reg}=7$ is the number of WiggleZ regions we use for the calculation of the combined  correlations. 
We do this for $\xi$, $\zeta$ and $Q$.
 In principle there should be no difference between calculating 
correlations using this method and measuring triplet counts across the whole survey; in practice our method
is more computationally  efficient and  gives us extra systematics tests by allowing us to compare results
region by region.

Overall diagonal errors and covariance matrices are calculated 
by jack-knife resampling the whole set of survey regions (see Figure \ref{fig:angsf}). 
 This way, we produce a catalogue of $N^{JK}_{tot}=88$ measurements. The variance of the correlations
 is calculated  as
\begin{equation}
\sigma_{X_{\rm{comb}}}^2 = \frac{N_{tot}^{JK}-1}{N_{tot}^{JK}}\sum_{j=1}^{N_{tot}^{JK}}(X_j^2 - \langle X\rangle^2)
\end{equation}
We also use this method to calculate the covariance matrix, which is used in the Maximum Likelihood approach to measure
the galaxy bias and cosmological parameters.

\subsection{The Combined WiggleZ 3PCF}

Figure \ref{fig:wz3pt0p6} shows the measurements of the redshift-space 2PCF $\xi(s_3(s,u,\theta))$,
 connected 3PCF $\zeta(s_1,s_2,\theta)$
 and  reduced 3PCF, $Q(s_1,s_2,\theta)$, of
 WiggleZ galaxies (from optimally combining the seven 
independent regions) in all redshift slices for a range of scales $s_1=10$, 15, 20 and 30 $h^{-1}$Mpc and shape 
$u=2$ as
a function of $\theta$. We have additionally measured the correlations using $u=1$ and $u=3$, for a total of
120 configurations, which are shown in Appendix A. 

We note qualitatively  that
we recover the same shape dependence of the galaxy 3PCF (mostly noticeable when looking at $Q$)
 which has been observed for galaxies at low redshift
\citep{mcbride_etal:11a, marin:11},
namely a bigger 3PCF amplitude
at small and large $\theta$, i.e the collapsed and elongated configurations.  
This `V-shape' is more prominent for large scales and elongated shapes; 
it is a  consequence of  the morphology of galaxy structures varying from spherically-shaped clusters and groups on 
small $\sim$ 1 $h^{-1}$Mpc scales to filaments on the largest  scales. This shape dependence of the 3PCF
depends on the galaxy type under investigation. It has been observed in SDSS \citep{kayo_etal:04,mcbride_etal:11a} 
and 2dFGRS \citep{gaztanaga_etal:05} that $L_*$ blue galaxies tend to have small 
3PCF amplitudes on small scales and very pronounced V-shapes on large
scales, compared to red galaxies and to $L>L_*$ galaxies, such as LRGs \citep{marin:11}. Although at a different
redshift, we find that the shape dependence of the WiggleZ 3PCF agrees with these lower redshift measurements.

We also note that on larger scales the behaviour of the reduced 3PCF $Q(\theta)$
for the most elongated shapes  is more
erratic, specially  for the $z=0.35$ and $z=0.68$ redshift slices. 
This is due to the fact that as $\xi \rightarrow 0$ on large scales, the measurements of $Q$ are less robust
and its errors become non-Gaussian.
But if we turn to analyze $\zeta$ instead, we can see that it is adequately measured up to the largest scales shown here, 
$s_3\sim100$ $h^{-1}$Mpc.

Comparing the clustering signal in different redshift slices, we can see that 
in general the differences in the  3PCF are small and  the signal is weaker
than in the case of the central $z_{\rm{eff}}=0.55$ slice.
This does not necessarily  indicate that there is no evolution of the clustering
of WiggleZ galaxies with redshift; the underlying dark matter clustering changes with redshift, and consequently 
the linear and non-linear bias factors evolve. From the 2PCF measurements we can estimate the evolution of the
linear bias, and using the 3PCF we can also test if there is evolution in the non-linear bias parameter.

\begin{figure}
\includegraphics[width=20pc]{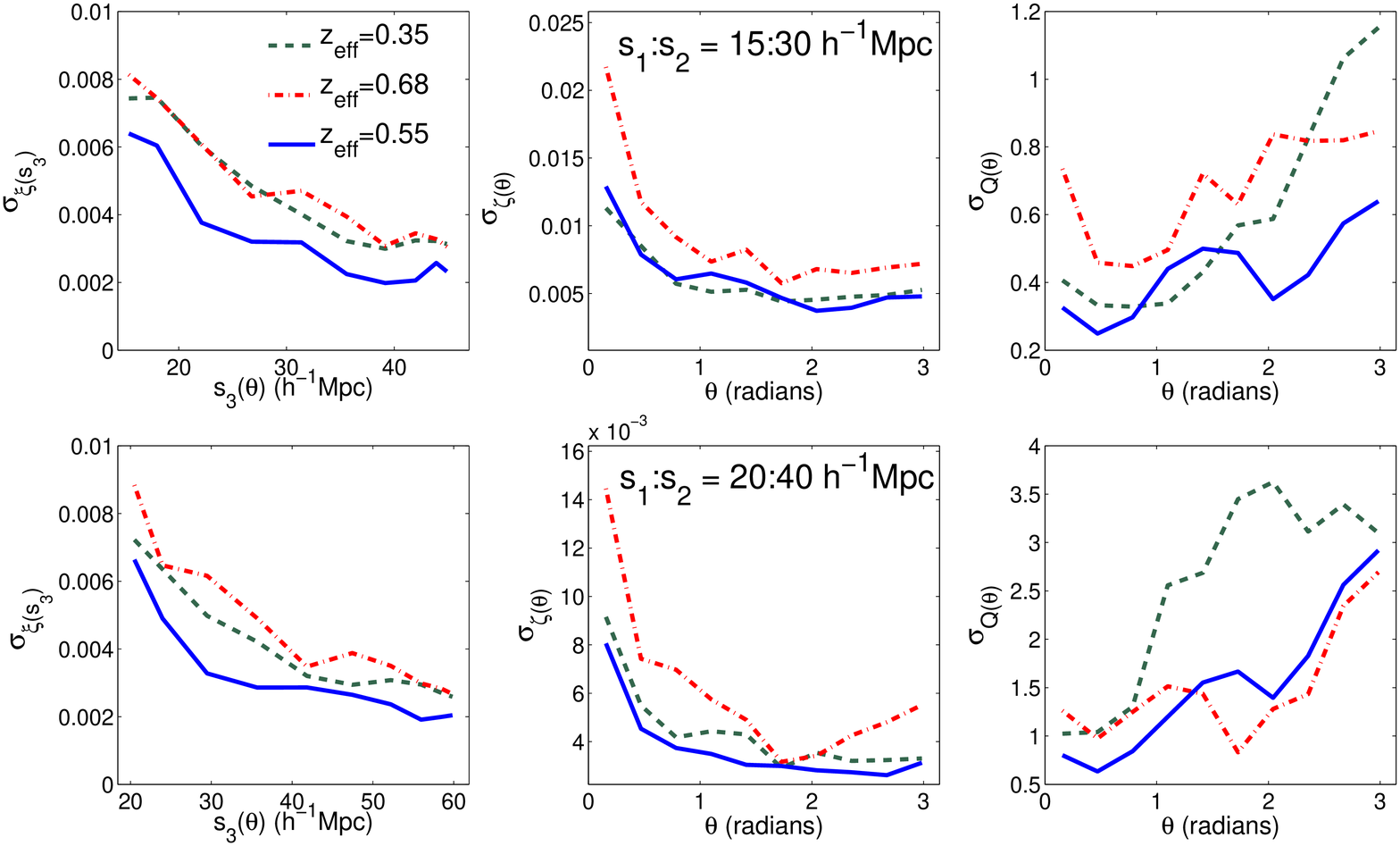}
\caption{\label{fig:jkdiags}
Diagonal errors of the correlation functions from jack-knife resampling. Top: errors for the
$s=15$ $h^{-1}$Mpc, $u=2$ triangles for the 2PCF $\sigma_\xi$ (left), connected 3PCF $\sigma_\zeta$ (centre),
and reduced 3PCF $\sigma_Q$ (right); solid line corresponds to measurements in the $z_{\rm{eff}}=0.55$ slice, 
dashed line for the $z=0.35$ slice, and dotted-dashed line for the $z=0.68$ redshift slice.
Bottom: same quantities for the $s=20$ $h^{-1}$Mpc, $u=2$ triangles.} 
\end{figure}

{\color{black} In Figure \ref{fig:jkdiags} we illustrate how the errors in our measurements vary with redshift by showing
 the 1-$\sigma$ diagonal errors (from jack-knife resampling) of the 2PCF and 3PCF  measurements for 
  a selection of configurations ($s=15,20$ $h^{-1}$Mpc and $u=2$) of  our (combined) redshift samples. 
 It can be clearly seen that measurements are more accurate in the central redshift slice  ($z_{\rm{eff}}=0.55$) than in the 
 outer ones. In these configurations, the relative error in the 2PCF is around  $\sigma_\xi/\xi \sim 0.15$, 
 for the connected 3PCF $\zeta$ the relative errors  reach $\sigma_\zeta/\zeta \sim 0.5$   in the central redshift
 slice.  }

\subsection{Covariance estimation}

 \begin{figure}
\includegraphics[width=21pc]{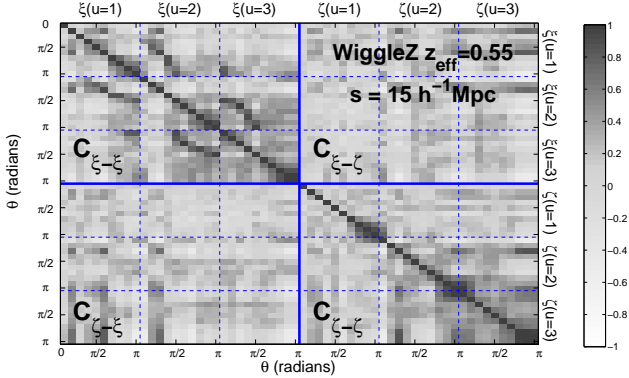}
\caption{\label{fig:s15cov}
The correlation matrix of the WiggleZ 2PCF and 3PCF of triangles with $s_1=15$ $h^{-1}$Mpc with 
$u=1,2$ and 3. Each element of the matrix is the covariance of each $s, u$ and $\theta$ triplet. }
\end{figure}
 
{\color{black} We  estimate the correlations between measurements of the 2PCF $\xi$ and 3PCF $\zeta$ by empirically
 calculating the covariance matrix. Using the jack-knife method, given a number of measurements $N_{JK}$ in number of bins
 $N_b$, a fractional
 error of a quantity $X$ for the sample $k$ can be written as
 \begin{equation}
 \Delta_i^k = \frac{X_i^k-\langle X_i \rangle}{\sigma_{X_i}}
 \end{equation}
 where in our case, if $i\leq N_b$ then $X=\xi(s_3)$ and otherwise $X=\zeta(s_1,s_2,s_3(\theta))$, and  $\sigma_{X_i}$ is the standard 
 error on $X_i$ calculated using the jack-knife method. Then we calculate the  correlation matrix (covariance matrix normalized by diagonal errors) as 
 \begin{equation}
 C_{ij}=\frac{1}{N_{JK}}\sum_{k=1}^{N_{JK}} \Delta_i^k \Delta_j^k
 \end{equation}
As an example that shows the observed behaviour of all configurations at different redshift samples, we show  
 in Figure \ref{fig:s15cov} the correlation matrix  
of both $\xi(s_3(\theta))$ and $\zeta(\theta)$ for the configurations with $s=15$ $h^{-1}$Mpc, and $u=1$, 2 and 3. In this
case the number of bins $N_b=30$, making $C_{ij}$ a $60\times60$ matrix.}
We divide this matrix into four regions depicting the auto- and cross- correlations.
 In the case of the $\xi(s_3)-\xi(s_3)$ covariance
 we notice that although it is dominated by diagonal terms,
  the off-diagonal terms are important too; the white stripes (signaling
high covariance) in the off-diagonal matrices ($\xi(u_1,s_3)-\xi(u_2,s_3)$ where $u_1\neq u_2$) are triangles
that share a similar $s_3$. The $\zeta$-correlation matrix also shows important non-diagonal elements that 
are more
correlated for the elongated shapes (a combination of true covariance and binning). 
The $\xi(s_3)-\zeta$ cross-covariance is small but needs to be considered in the analysis.

\section{Constraints on Galaxy bias and $\sigma_8$}
\label{sec:constraints}

We  compare the dark matter correlations measured in  the GiggleZ simulations to the WiggleZ
 2PCF and 3PCF in order to constrain the linear and non-linear bias parameters using the local bias model
 described in $\S$3.3.
In this analysis we assume all cosmological parameters are fixed, and fit for $\sigma_8(z)$ by scaling
the amplitude of the dark matter correlations in the manner of equations (7) and (8), 
with all quantities measured in redshift space.
{\color{black} We want to emphasize that  since we are using N-body simulation measurements to compare to our galaxy correlations,  and not 
an analytical model, the only cosmological parameter we can modify is $\sigma_8(z)$ through the ratio $\sigma_8/\sigma_{\rm{8,fid}}$, where
the fiducial value corresponds to the one used in the GiggleZ simulation $\sigma_{8,{\rm fid}}(z=0)=0.812$.}

\subsection{Methods}

We carry out a Maximum Likelihood parameter estimation test, where 
we look to minimize the quantity
\begin{equation}
\chi^2=\sum_{i=1}^{i=2N_b}\sum_{j=1}^{j=2N_b}\Delta_iC_{ij,SVD}^{-1}\Delta_j
\label{eq:chi2}
\end{equation}
where $N_b$ is the number of triangular configurations used; 
we have $N_b$ distances where we measure the 2PCF of $s_3$, and $N_b$ triangles where we 
measure the 3PCF, therefore we have $(2N_b)^2$ elements in our covariance matrix. The value
of $\Delta_i$ is the difference between the correlation measured and the biased DM correlation: 
\begin{eqnarray}
\Delta_i &=& (\xi(s_3)^{obs}_i - \xi(s_3)^{model}_i)/\sigma_{\xi(s_3)i}, \textrm{ for } i\leq N_b\\
\Delta_i &=& (\zeta(s,u,\theta)^{obs}_i-\zeta(s,u,\theta)^{model}_i)/\sigma_{\zeta(i)}, \textrm{ for } i>N_b
\end{eqnarray}
where $\xi^{model}$ and $\zeta^{model}$ are given by eqs. (7,8).

In order to invert the covariance matrix we use the approach 
of \cite{gaztanaga_scoccimarro:05} and repeated in several 
3PCF works \citep{gaztanaga_etal:05,mcbride_etal:11a, marin:11},
 of employing only the highest eigenmodes of the covariance matrix to minimize effects of numerical noise. 
  We employ the Singular Value Decomposition method, where our normalized covariance
 matrix can be decomposed  $C=UDV^T$ {\color{black} (and where $V=U$ for a symmetric matrix)}, 
 where the diagonal matrix $D_{ij}=\lambda^2_i\delta_{ij}$ stores
 the eigenvalues {in decreasing order, \color{black} and the columns of the matrix $U$ stores the eigenmodes of $C$} . 
 When inverting the matrix $C^{-1}=VD^{-1}U^T$, {\color{black}(where $D_{ij}^{-1}=(1/\lambda_i^2)\delta_{ij}$)
 we need to discard some of these eigenmodes, meaning set some $D_{ii}^{-1}\equiv 0$.}
   Firstly, we use a finite number of jack-knife samples to estimate our covariance.
  Since using jack-knife samples assumes the statistical independence of the subsamples, 
  the jack-knife regions {\color{black} should have a larger spatial extent}
   than the largest
 scales studied.  {\color{black} In our case we use $N_{JK}=88$ JK regions
 because our largest scales are of the order $\sim 100$ $h^{-1}$Mpc. 
 However, we are using a large number of 2PCF and 3PCF measurements, and
generally  $N_{JK}<N_b$, which will make our matrix singular for modes $i>N_{JK}-1$ (see \citealt{press_etal:92}  for instance).
Therefore, all those eigenmodes have to be discarded.}
  A second cut comes from the fact that {\color{black} even though the covariance matrix is not singular when the first cut is applied, there 
  are still eigenvalues $\lambda_i^2$ with very low numerical value, which will make unstable the inversion of
  $C_{ij}$ with $\lambda_i^2 < \sqrt{2/N_{JK}}$ \citep{gaztanaga_scoccimarro:05}. Adding these
 unstable modes, as explained by these authors and later in section 5.2, is equivalent to introducing artificial `signal' to our measurements 
 that  will bias our fits.}

\begin{figure}
  \epsfig{file=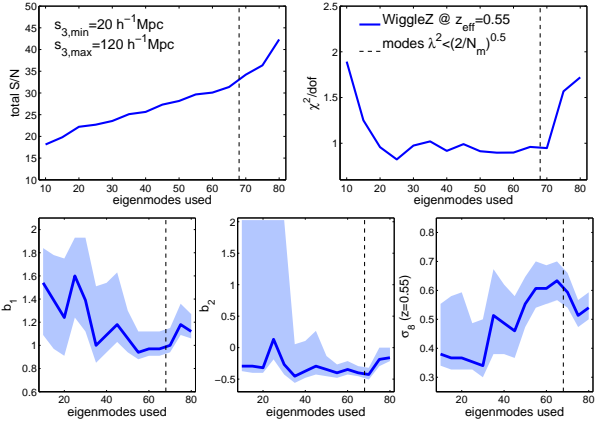,width=8cm}\\
  \caption{\label{fig:compsev} {\color{black}Dependence of the best-fitting bias parameters and $\sigma_8$ on the number of eigenmodes 
 used for the $z_{\rm eff}=0.55$ 2PCF and 3PCF analysis. 
 In the left panel we show eigenmodes used after SVD (from the  $\lambda_i^2> \sqrt{2/N_{JK}}$
 limit), with fixed  $s_{3,{\rm min}}=20$  and  $s_{3,{\rm max}}=120$ $h^{-1}$Mpc. The top left panel shows the total $S/N$ according to 
 equation (\ref{eq:sn}) as a function of number of eigenmodes used. The top right panel shows the $\chi^2/$dof, and lower panels
 show the 68\% CL intervals for $b_1$, $b_2$ and $\sigma_8$. The vertical dashed line 
 represents the eigenmode limit where  $\lambda_i^2 = \sqrt{2/N_{JK}}$, with $N_{JK}=88$.}}
  \end{figure}

\begin{figure}
  \epsfig{file=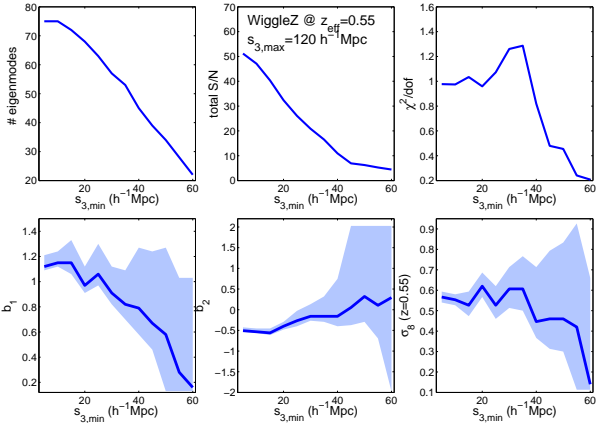,width=8cm}\\
 \caption{\label{fig:compsrmin} {\color{black} Dependence of the best-fitting bias parameters and $\sigma_8$ on the fitting range $s_{3,{\rm min}}$, 
 with  $s_{3,{\rm max}}=120$ $h^{-1}$Mpc for the  $z_{\rm eff}=0.55$ 2PCF and 3PCF analysis. The top left panel shows the number of eigenmodes
 used where $\lambda_i^2 = \sqrt{2/N_{JK}}$ as a function of  $s_{3,{\rm min}}$. The top middle panel shows the total $S/N$, top right panel 
 the $\chi^2/$dof of the best fit parameters. Lower panels
 show the 68\% CL intervals for $b_1$, $b_2$ and $\sigma_8$. }}
 \end{figure}
 
 \begin{figure}
  \epsfig{file=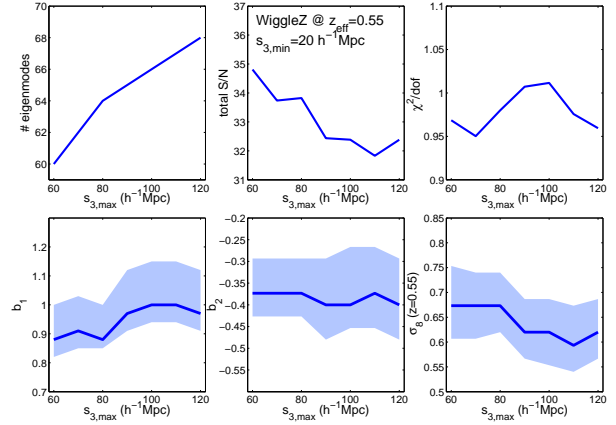,width=8cm}\\
 \caption{\label{fig:compsrmax}Same as Figure 11, in this case fixing $s_{3,{\rm min}}$=20 $h^{-1}$Mpc  and having quantities shown as a function
 of $s_{3,{\rm max}}$.}
 \end{figure}

 We also have to set the minimum and maximum scale of the model fit. 
 {\color{black} In our analysis this means that we chose configurations with the third size in a range $s_3=[s_{3,{\rm min}},s_{3,{\rm max}}]$.
The minimum scale is given by the range of validity of the local bias model.
 The maximum scale could be set by systematics in the selection function or when
  the correlation signal is weak.}
  {\color{black} In Figures \ref{fig:compsev}, \ref{fig:compsrmin} and \ref{fig:compsrmax} we show how our choices of the number of eigenvalues
  used, the values of  $s_{3,{\rm min}}$ and $s_{3,{\rm max}}$ respectively  affect our constraints for our $z_{{\rm eff}}=0.55$ sample.}
  
{\color{black} We wish to make a sensible default choice for these
  options and establish that our essential conclusions are not very
  sensitive to this choice.  Figure \ref{fig:compsev} investigates the
  dependence of our results on the number of eigenmodes included,
  fixing $s_{3,{\rm min}}=20$ $h^{-1}$Mpc and $s_{3,{\rm max}}=120$
  $h^{-1}$Mpc.  We observe that if we use a small number of
  eigenvalues we have large variations in the best fit values and poor
  constraints, especially in $b_2$; then there is a range when the
  number of eigenvalues used is $\sim 50$ where our results are
  insensitive, and when we include modes with $\lambda_i^2 >
  \sqrt{2/N_{JK}}\sim 65$ we again have unstable behaviour.  In this
  figure we also consider the dependence of the minimum $\chi^2$ per
  degree of freedom on the number of eigenmodes employed, where the
  degrees of freedom are equal to the eigenmodes used minus the number
  of parameters we seek to constrain.  We note that good fits are
  produced for a wide range of choices.  We can also estimate the
  total signal-to-noise ($S/N$) ratio of the modes used as
  \citep{gaztanaga_scoccimarro:05}:}
\begin{equation}
\label{eq:sn}
\left(\frac{S}{N}\right)_i=\frac{1}{\lambda_i}\sum_{j=1}^{j=2N_b}U_{ji}\frac{X_j}{\sigma_{X_j}},
\end{equation}
where $X_j=\xi_j$ when $j<N_b$, and $X_j=\zeta_{j-N_b}$ when $j>N_b$.

{\color{black} Our conclusion from this analysis is that any systematic
  fluctuations in our parameter measurements do not dominate the
  statistical errors for a wide range of choices.  Our default fits
  are performed for an eigenmode cut $\lambda_i^2 = \sqrt{2/N_{JK}}$
  and a fitting range $s_{3,{\rm min}}=20$ $h^{-1}$Mpc and $s_{3,{\rm
      max}}=120$ $h^{-1}$Mpc for the $z_{\rm{eff}}=0.55$ and 0.68
  slices, and $s_{3,{\rm max}}=100$ $h^{-1}$Mpc for the
  $z_{\rm{eff}}=0.35$ sample.}

Another aspect to consider in our analysis is that the measured redshifts in our survey contain a small
fraction of  `redshift blunders' \citep{blake_etal:10sf},
 failures that tend to wash out the clustering we measure
in our galaxy field. This redshift blunder fraction is $f_b=0.03$ for the $z_{\rm{eff}}=0.35$ and 0.55 redshift slices, and
$f_b=0.05$ for the outer $z_{\rm{eff}}=0.68$ slice; the correction to the 2PCF 
(since $\xi\propto\langle\delta_{\rm{gal}}\delta_{\rm{gal}}\rangle$)
 is to multiply the data and errors by
 $(1 - f_b)^{-2}$, meaning that the clustering amplitude increases; and the correction to the 3-pt function (where
 $\zeta\propto\langle\delta_{\rm{gal}}\delta_{\rm{gal}}\delta_{\rm{gal}}\rangle$) is $(1 - f_b)^{-3}$.

\subsection{Constraints at $z_{\rm{eff}}=0.55$}

 \begin{figure*}
 \centering
 \begin{tabular}{cc}
 \epsfig{file=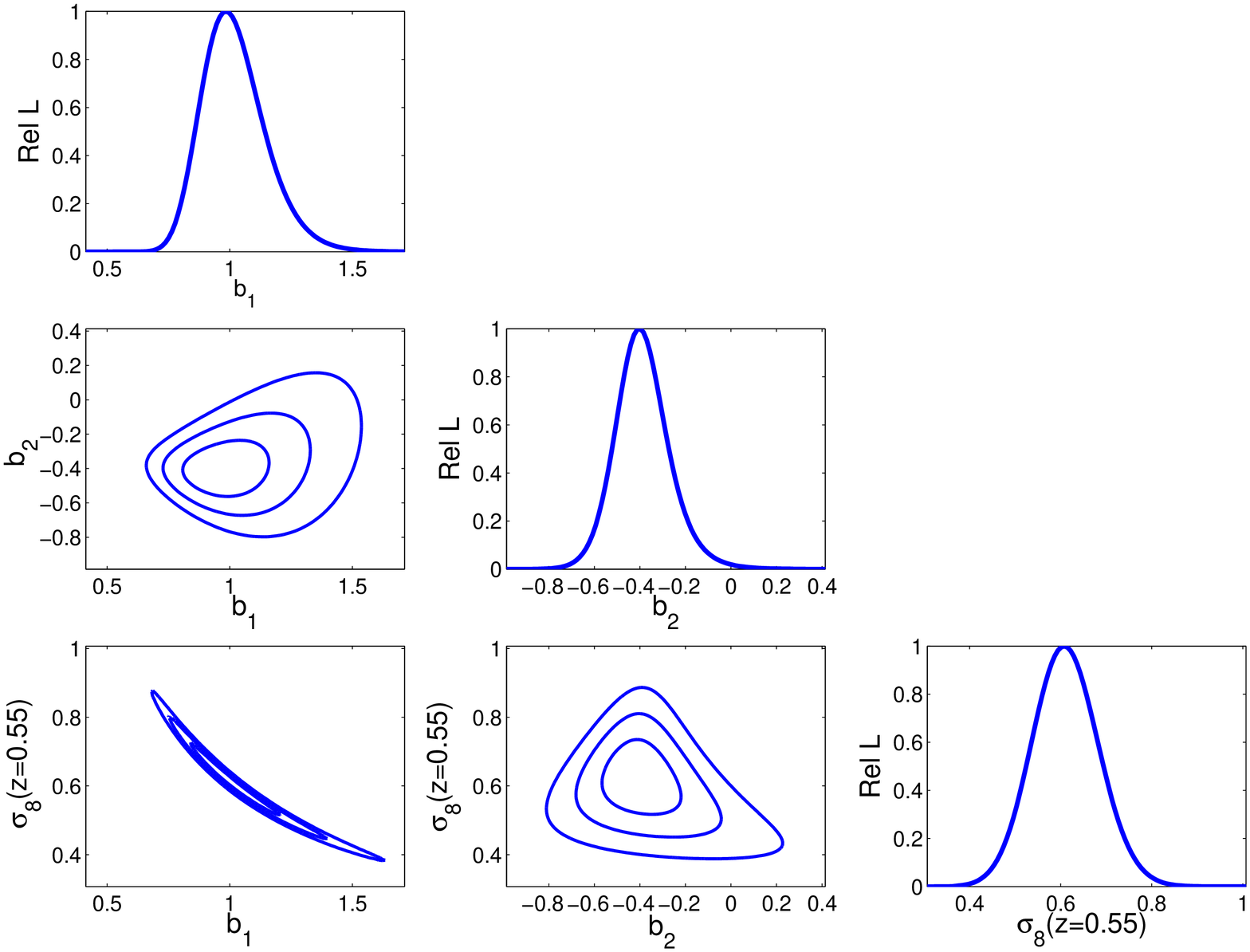,width=8cm}&
 \epsfig{file=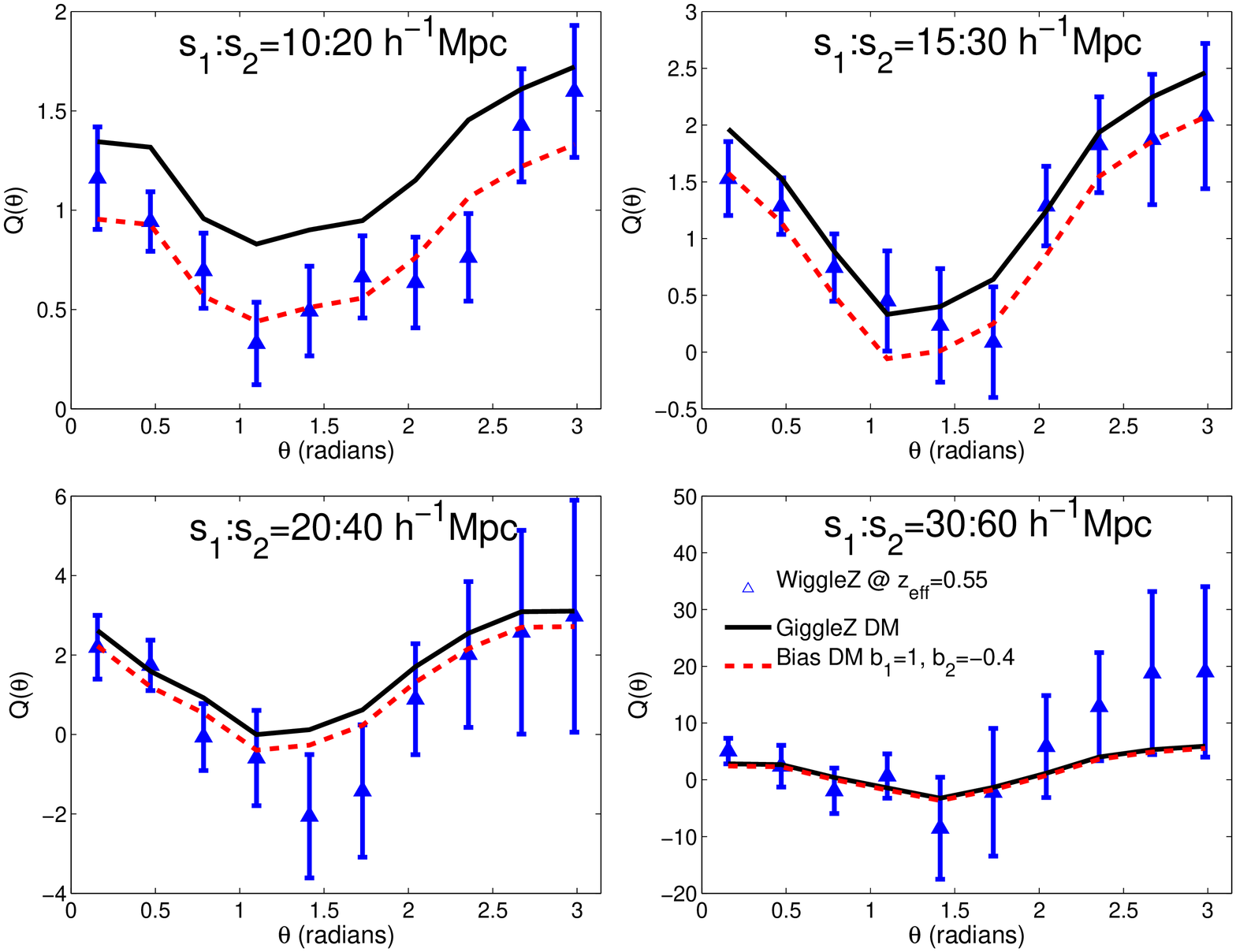,width=8cm}
 \end{tabular}
 \caption{\label{fig:b1b2s80p55} Left: Constraints on the bias parameters $b_1$, $b_2$ and $\sigma_8$ for the 
 $z_{\rm{eff}}=0.55$ WiggleZ redshift slice. The contours represent 1-$\sigma$, 2-$\sigma$ and 3-$\sigma$ joint confidence regions for a 
 two-parameter fit. Right: Dark Matter reduced 3PCF (black thick line), the WiggleZ $Q(\theta)$ for the $z_{\rm{eff}}=0.55$ slice
 and the biased dark matter $Q(\theta)$ (dashed line) using the best-fit parameters found in this analysis.}
 \end{figure*}

In Figure \ref{fig:b1b2s80p55} we show the measurements of $b_1$, $b_2$ and $\sigma_8$ in the
 $z_{\rm{eff}}=0.55$ redshift slice. 
 The  measured linear bias, 
$b_1 \sim 1$,  agrees with  values obtained for WiggleZ galaxies using other methods
\citep{blake_etal:11f, contreras_etal:12w}, using 2-point statistics,
where in our study  we additionally marginalise over all values
of $\sigma_8$.  We detect a significantly non-zero value for the non-linear bias $b_2\sim -0.4$.

We measure the amplitude of fluctuations $\sigma_8(z)$ with 10\% accuracy and find that our results agree with 
independent predictions, based on cosmological parameter measurements from the CMB in a $\Lambda$CDM model.
It is important to note that these estimates are independent of any other observable than the galaxy clustering itself. 
Extrapolated to $z=0$, our measurements of $\sigma_8$ from WiggleZ galaxies are 
 consistent with conclusions from 3PCF measurements of other tracers such as the LRGs \citep{marin:11},
 which provides evidence that this method is robust against the type of galaxy used. 

As is shown in Table \ref{tab:c1c2s8}, the empirical bias model is an adequate fit to the data, and 
that can be graphically seen in the right-hand plots of Figure \ref{fig:b1b2s80p55}, where we show the biased 
dark matter $Q(\theta)$, which in our model depends on the bias parameters but not on $\sigma_8$:
\begin{equation}
Q_{\rm{gal}}=\frac{1}{b_1}\left(Q_{\rm{m}}+\frac{b_2}{b_1}\right).~
\end{equation}
We can see that our galaxy bias model is adequate on all scales, but our fits are 
 mainly driven by the $s=10$ and 15 $h^{-1}$Mpc 
configurations, which have the highest signal-to-noise ratio.

 \begin{table}
\begin{minipage}{80mm}
\centering
 \caption{\label{tab:c1c2s8}Constraints on bias parameters and $\sigma_8$ for WiggleZ samples}
  \begin{tabular}{l c c c c c}
  \hline
    $z_{\rm{eff}}$ &$b_1$&$b_2$&$\sigma_8$&$\chi^2/$dof&$S/N$\\
 \hline
 0.35 & $0.72_{-0.14}^{+0.14}$ & $-0.36_{-0.08}^{+0.11}$&$0.69_{-0.11}^{+0.12}$&1.10&3.25\\
 0.55 & $0.99_{-0.09}^{+0.10}$ &$-0.41_{-0.08}^{+0.09}$ &$0.61_{-0.05}^{+0.06}$ &0.96&4.99\\
 0.68 & $1.06_{-0.18}^{+0.16}$ & $-0.48_{-0.12}^{+0.14}$ &$0.53_{-0.07}^{+0.08}$ &0.82&4.62\\
 \hline
\end{tabular}
\end{minipage}
\end{table}

\subsection{Constraints at different redshifts}

\begin{figure}
\begin{center}
\includegraphics[height=15pc,width=22pc]{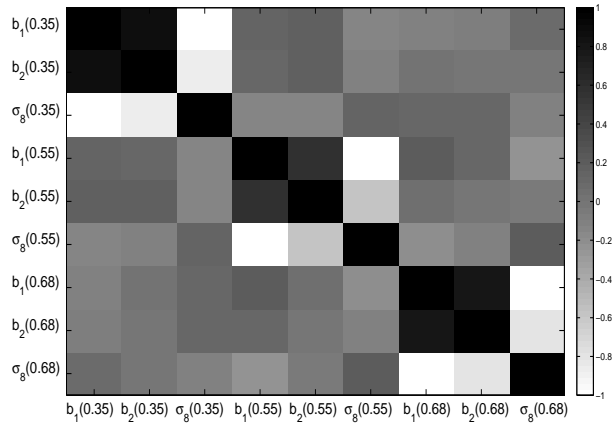}
\end{center}
\caption{\label{fig:covbs8}Correlation matrix of  WiggleZ best fit parameters $b_1$, $b_2$ and $\sigma_8$ as function of redshift}
\end{figure}

We repeated the  analysis of the correlation functions for the other two redshift slices in order to get
constraints on the bias parameters and $\sigma_8$ as a function of redshift. In Table \ref{tab:c1c2s8} we show
our results. In general we can see that there is clear trend in all the parameters with redshift,
and that from the values of $\chi^2/$dof,  our model of the bias provides a good fit. 
{\color{black} We also estimated the covariance between the best-fit parameters in different redshift slices, expected due to the
overlap between redshift samples.  In order to make this measurement we fitted the bias parameters and $\sigma_8$ to each delete-1 jack-knife sample,
obtaining $N_{JK}=88$ sets of best fit parameters which we used to construct a covariance matrix. We show the correlation matrix of these parameters in Figure \ref{fig:covbs8}. We 
observe that in any individual redshift slice there is a positive correlation between $b_1$ and $b_2$ and a negative correlation between these parameters and
 $\sigma_8$. Although there is overlap in the redshifts of the samples studied, the correlation coefficients between best-fit  parameters in separate redshift slices are small.}

 In the following we study the change  of the bias parameters and $\sigma_8(z)$ with redshift in more detail.

\begin{figure*}
 \centering
 \begin{tabular}{cc}
 \epsfig{file=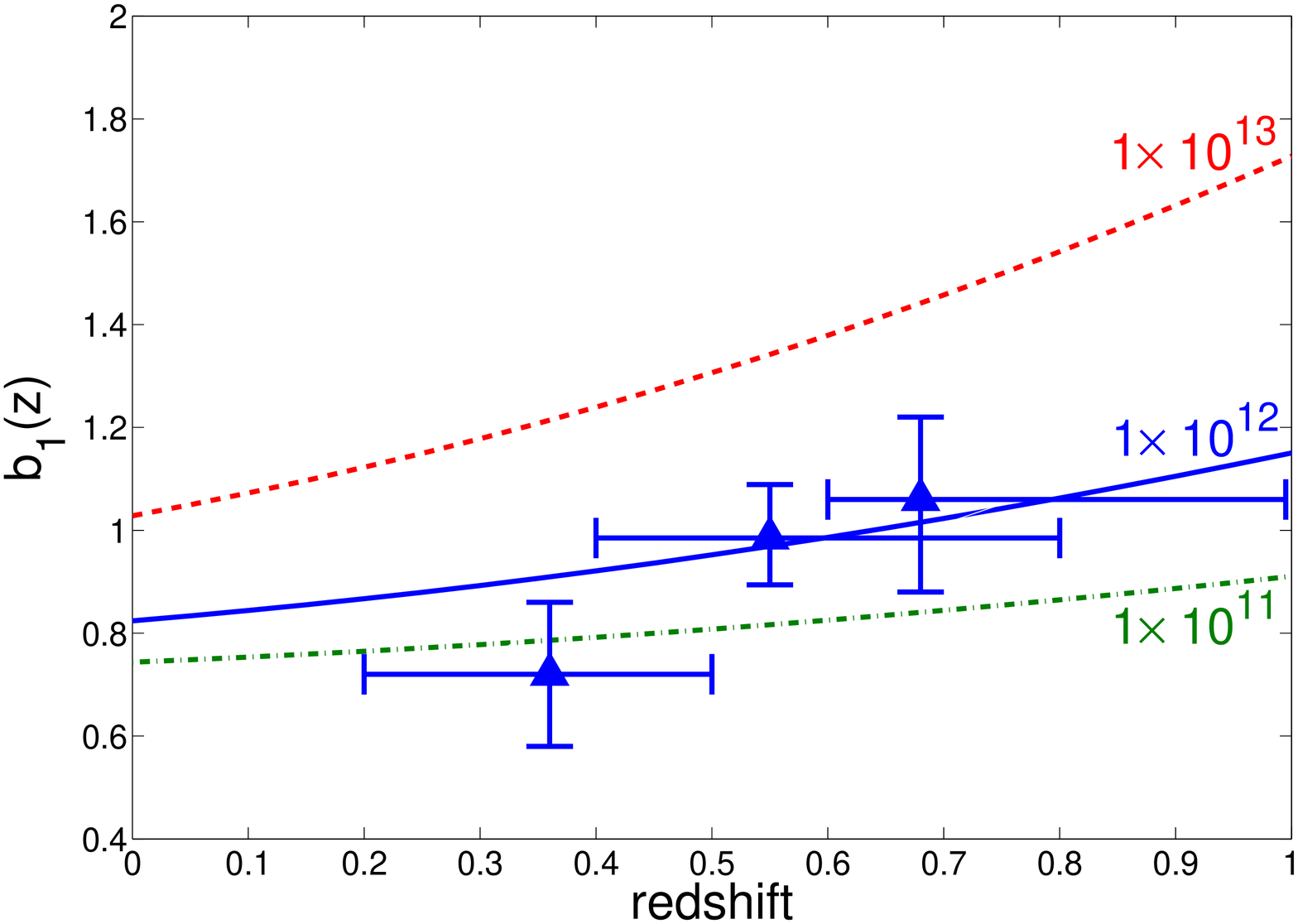,width=9cm}&
 \epsfig{file=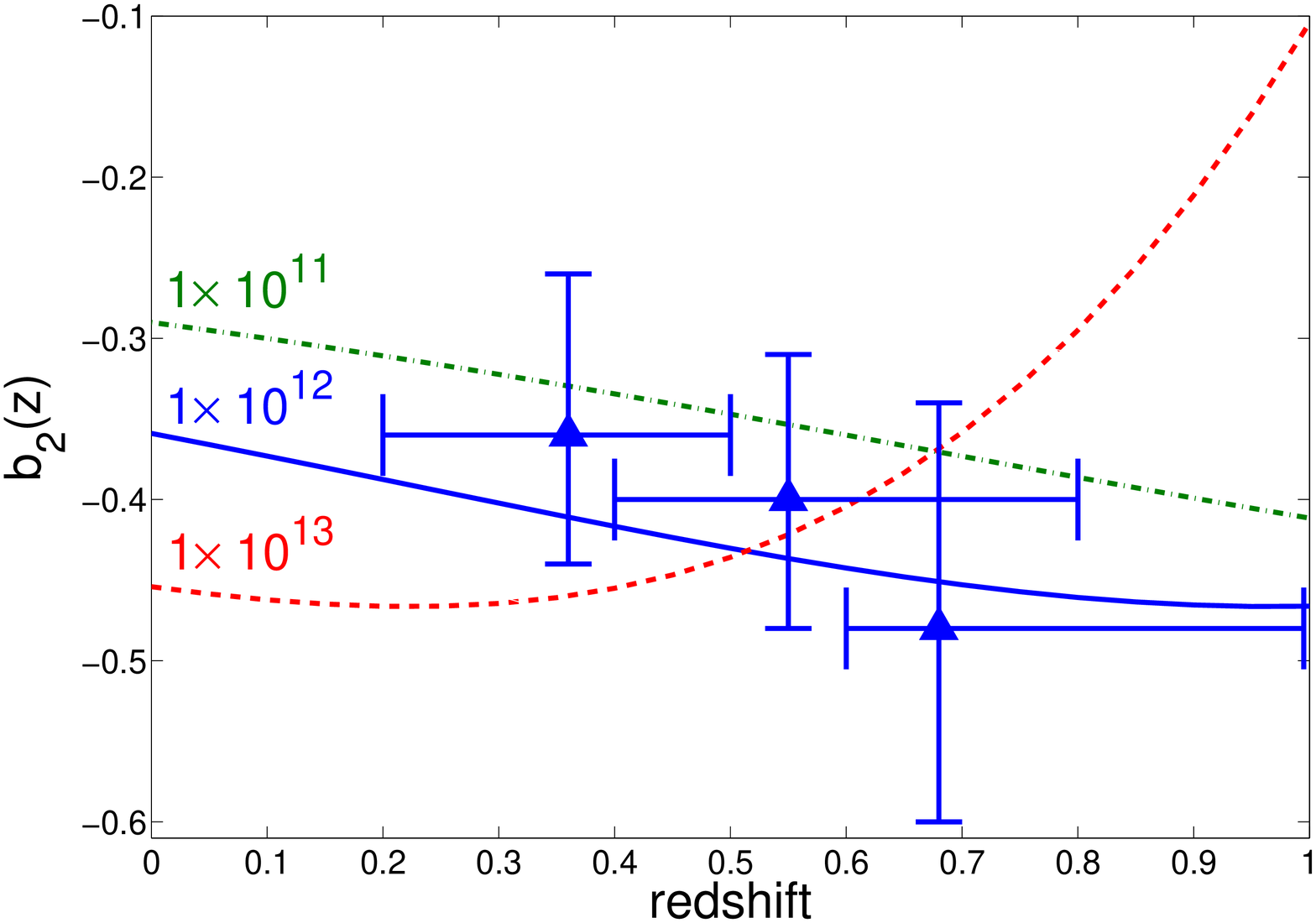,width=9cm}
 \end{tabular}
 \caption{\label{fig:b1b2evol} Evolution of the bias parameters. Left: Evolution of linear bias: triangles are best
 fit parameters from our WiggleZ regions. Lines are halo model prediction of bias for halos of masses
 $M_h=10^{11}$ $h^{-1}M_\odot$ (green dashed-dotted line),  $M_h=10^{12}$ $h^{-1}M_\odot$ (blue solid line) 
 and $M_h=10^{13}$ $h^{-1}M_\odot$ (red dashed line).Right: evolution of the non-linear bias parameters.}
 \end{figure*}

\subsubsection{Constraints of WiggleZ bias}

Figure \ref{fig:b1b2evol} displays the change  of the galaxy bias with redshift.
{\color{black} Due to our selection criteria, we are not necessarily selecting the same population of galaxies at different redshifts
\citep{li_etal:12}, specially at redshifts $z<0.5$. Therefore we only can make general statements about evolution of 
the bias of WiggleZ galaxies.}
For the linear bias  $b_1$ it can be seen
that there is an upward trend of bias with redshift, consistent with what is expected for the evolution of halos of a given mass: 
 massive objects are less
common in older times, and therefore more biased.

In order to compare our measured bias evolution to the expectation of simple  models, we also plot the 
evolution of the bias of dark matter halos of fixed mass with redshift in the halo model. 
These are given by the analytical expectation using Sheth-Tormen mass functions \citep{sheth_etal:01} for the
dark matter halos, and the linear and non-linear bias from the work of  \cite{scoccimarro_etal:01}.
Since our galaxies are a subsample of the total population, with a particular colour and luminosity selection,
we do not expect that they should follow exactly one track of evolution, but in any case, our measured bias evolution
seems to agree with those galaxies living in halos with masses $\sim 10^{12}$  $h^{-1}M_\odot$.

In our measurements and in the halo model, it is expected that when a galaxy tracer
has a linear bias $\sim 1$, it should have a small but significantly negative non-linear bias $b_2$. In the right panel 
of Figure  \ref{fig:b1b2evol} 
we show the evolution of the WiggleZ galaxies' non-linear bias. These have negative values, 
and  their trend agrees with what is expected of $\sim 10^{12}$  $h^{-1}M_\odot$ halos.
A more  detailed analysis
using Halo Occupation Distribution models is needed to have a complete picture of how WiggleZ galaxies
populate dark matter halos; this is beyond the scope of this paper.

\subsubsection{Evolution of cosmic growth}

\begin{figure}
\begin{center}
\includegraphics[height=15pc,width=22pc]{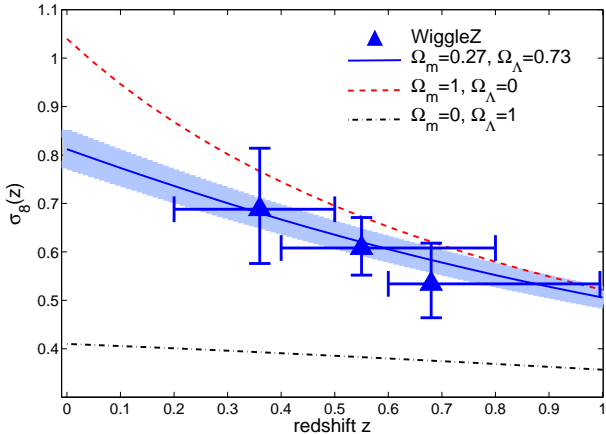}
\end{center}
\caption{\label{fig:sig8z}Evolution of $\sigma_8(z)$. Blue symbols correspond to the estimates of $\sigma_8(z_{\rm{eff}})$
 from the 2PCF and 3PCF of WiggleZ galaxies at different redshifts, marginalizing over the linear and non-linear bias parameters.
The blue solid line  corresponds to the evolution of  $\sigma_8(z)$ in a flat WMAP5-$\Lambda$CDM universe, with $\Omega_{\rm{m}}=0.27$ and $\sigma_8(z=0)=0.812$; the blue shaded region corresponds to combined WMAP5 errors.  
The red dashed line corresponds to the evolution of
  $\sigma_8(z)$ in a flat $\Omega_m=1$ CDM universe. The black dash-dotted line shows the evolution of a flat $\Omega_\Lambda=1$ universe.
  All models are normalized at the epoch of recombination.}
\end{figure}

In Figure \ref{fig:sig8z} we plot our measurements of $\sigma_8$ as a function of redshift from the WiggleZ survey data.
 In linear theory, the value of $\sigma_8$ is calculated as 
 \begin{equation}
\sigma^2(R=8,z)= \int \frac{d^3k}{(2\pi)^3} \left| W(k,R=8) \right|^2P_{lin}(k,z)
\end{equation}
 where $W(k,R)$ is the Fourier transform of a tophat window of radius $R=8$ $h^{-1}$Mpc.  
The linear power spectrum evolves as $P_{lin}(k,z)\propto [D(z)/D(z_*)]^2P(k,z_*)$, where $z_*$ is a reference 
redshift (e.g., the redshift of recombination) and $D(z)$ is the linear growth factor, obtained from the solution
to the linearized equations of motion of primordial overdensities \citep{peebles:80, bernardeau_etal:02}. The evolution
of the linear growth factor depends on the parameters of the cosmological model \citep{lahav_suto:03}. 
Thus, we obtain
\begin{equation}
\sigma_8(z)=\frac{D(z)}{D(z=0)}\sigma_8(z=0).
\end{equation}
Therefore, $\sigma_8(z)$ measurements from the 2PCF and 3PCF can be used to  study the evolution 
of the linear growth factor.

As predicted by the standard cosmological model,  the value of $\sigma_8(z)$ we measure  decreases at earlier times, in agreement
with the WMAP5 cosmological parameters. 
{\color{black} Assuming a flat $\Lambda$CDM model with $\Omega_m=0.27$ we find that when extrapolated to the present epoch, 
$\sigma_8(z=0)=0.79^{+0.06}_{-0.07}$.}
 Our results also agree with the latest estimation of $\sigma_8(z)$ using BOSS/SDSS-LRGs passive galaxies from  
\cite{tojeiro_etal:12}; modelling the evolution of the linear bias for their particular population they
 find similar values to ours. However, in our work we need to make no assumptions about the evolution of the bias, just
 in the validity (range of scales) of the empirical bias model we adopt. We also find agreement with other measurements of $\sigma_8$ from the 3PCF
 of the SDSS LRG sample  \citep{marin:11}; they find $\sigma_8(z=0.35)=0.65^{+0.02}_{-0.05}$, consistent with our measurements at the same 
 effective redshift.
 
  We also plot in  Figure  \ref{fig:sig8z}  the evolution of $\sigma_8$ in two different models of flat universes, an Einstein-de Sitter model
  (flat $\Omega_m$=1) and another one with no matter content, just cosmological constant. As WMAP5  \citep{komatsu_etal:09} measured 
  the amplitude of CMB perturbations at the epoch of recombination $z\sim 1100$, we normalize $\sigma_8$ at that redshift for the three
  cosmologies shown here. We find that an Einstein-de Sitter universe is disfavoured {\color{black} ($\Delta\chi^2=3.5$ for one parameter fit) 
  when combining measurements of the three WiggleZ redshift samples, 
  as well as spacially flat models with low matter content ($\Delta\chi^2=45.01$ for $\Omega_m=0.01$).}

\section{Summary and Conclusions}

We have measured the redshift space two- and three- point clustering statistics for the WiggleZ galaxies and obtained 
constraints on the linear and non-linear galaxy bias and the cosmological parameter $\sigma_8$ at three different epochs. 
Our results can be summarized as follows:

\begin{itemize}
\item We obtain significant measurements of the WiggleZ 3PCF, recovering its shape dependence on large
scales, spanning a wide redshift range for all regions and subregions (in angle and redshift) of the galaxy sample.
 \item These measurements are in agreement with standard models for the growth of structure driven by gravitational clustering,
 reflecting the morphology of the clustering large-scale structures, i.e. the `cosmic web'. 
\item Using a simple local bias parameterization along with an empirical treatment of redshift-space distortions of the correlation functions,
 we get constraints on the bias parameters as a function of redshift. Our
estimation of the linear bias agrees with  evolution of dark matter halos $\sim 10^{12}$ $h^{-1}M_\odot$.
\item For all our redshift samples, we detect a significant non-zero (negative) non-linear bias, also consistent with the models for the
non-linear bias evolution of  halos of masses  $\sim 10^{12}$ $h^{-1}M_\odot$.
\item We also constrain the  evolution of $\sigma_8$ with redshift, and by extension, the evolution of the linear growth factor. We find that
our measurements are  consistent
with the predictions of a WMAP5 $\Lambda$CDM concordance cosmology and with measurements from other methods and observables.

\end{itemize}

The improvement in the measurements of the higher-order correlations in the last ten years has been dramatic,  and 
it is remarkable that we can now measure 
the 3PCF using galaxies up to redshift $z\sim 1$. Although the signal-to-noise ratio of the WiggleZ 3PCF is weaker
 than lower-redshift measurements from the SDSS Main Sample 
(\citealt{mcbride_etal:11a}; \citealt{mcbride_etal:11b} ) and SDSS LRG sample \citep{kulkarni_etal:07, marin:11}, we nonetheless
have extended the utility of higher-order correlations functions to $z\sim1$, using the WiggleZ survey data. 
We note that using jack-knife resampling for error
estimation probably overestimates the variance on large scales \citep{marin:11}; our measurements would be improved by the availability
of mock galaxy catalogues. At the same time, 
with improved statistics we need to improve our modelling 
of redshift space distortions and small-scale effects
in order to extract as much information as possible from the higher-order correlations.

Also, improved modelling would help to combine 3PCF measurements with other observables such as 
 clustering and lensing \citep[see][]{mandelbaum_etal:12},  2-dimensional RSD 2-point statistics, or  the
  passive galaxies method \citep{tojeiro_etal:12}.

In the near future, with improved measurement techniques and with bigger surveys, 
we will be able to use these techniques to measure the growth factor accurately and discriminate between 
 $\Lambda$CDM model and  modified gravity models \citep{linder_cahn:07}.

 \section*{Acknowledgements}

We thank Eyal Kazin for fruitful discussions and suggestions, and to the anonymous referee and for valuable comments and
suggestions. 
We acknowledge financial support from the Australian Research Council
through Discovery Project grants which have funded the positions of
MP, GP, TD and FM.  SMC acknowledges the support of the Australian
Research Council through a QEII Fellowship. 
CB acknowledges the financial support of the ARC through a Future Fellowship award.
We are also grateful for support from the
 Centre for All-sky Astrophysics, an Australian Research Council Centre of Excellence
  funded by grant CE11000102.

GALEX (the Galaxy Evolution Explorer) is a NASA Small Explorer,
launched in April 2003.  We gratefully acknowledge NASA's support for
construction, operation and science analysis for the GALEX mission,
developed in co-operation with the Centre National d'Etudes Spatiales
of France and the Korean Ministry of Science and Technology.

Finally, the WiggleZ survey would not be possible without the
dedicated work of the staff of the Australian Astronomical Observatory
in the development and support of the AAOmega spectrograph, and the
running of the AAT.

\bibliography{wz3pt}

\appendix

\section{WiggleZ 3PCF measurements of isosceles and very elongated triangles}

\begin{figure*}
\includegraphics[width=12cm]{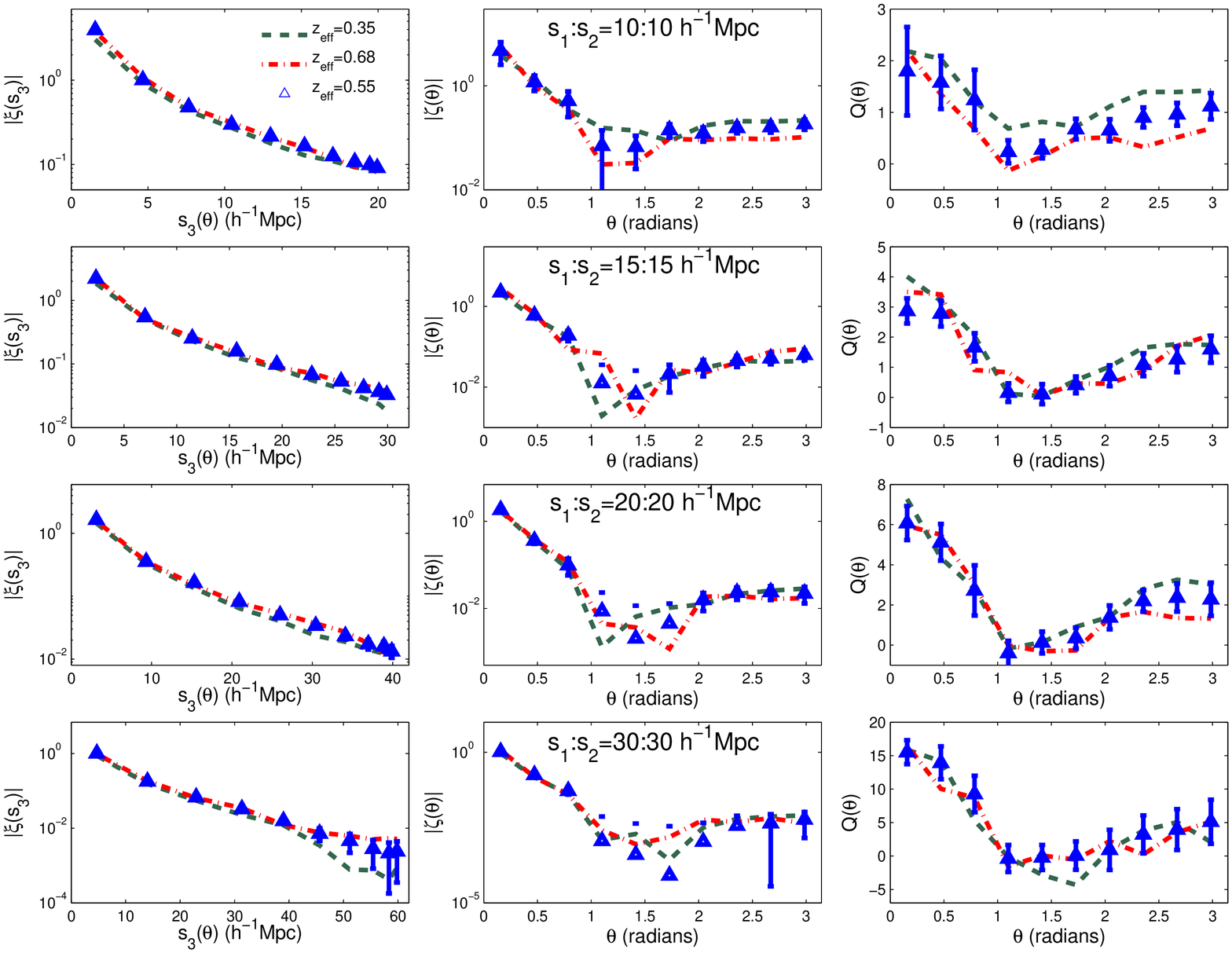}
\caption{\label{fig:wzu1}
The WiggleZ combined correlations of isosceles triangles ($u=1$) 
in redshift-space. We show the absolute 2PCF $|\xi(s_3(\theta))|$ (left), absolute connected 3PCF $|\zeta(s,u=1,\theta)|$ (middle), and 
reduced 3PCF $Q(s,u=1,\theta)$ (right) of WiggleZ galaxies in the $z_{\rm{eff}}=0.55$ slice (blue triangles),  
in the slice at $z_{\rm{eff}}=0.35$ (green dashed line) and in the $z_{\rm{eff}}=0.68$ slice (red dashed-dotted line).
 Errors have been determined by jack-knife resampling.}
\end{figure*}

In Figure \ref{fig:wzu1} we plot results of the WiggleZ correlation functions measured for isosceles configurations, 
where $u=s_2/s_1=1$, for the three redshift slices we studied. In these configurations, the third side length runs from
$s_3 \sim 0$ on very small angles $\theta \sim 0$, where the bias model we use is no longer valid due to high nonlinearities. 
Since the third triangle side $s_3$ covers a  large 
range of scales, for purposes of plotting we show the absolute values of the 2PCF and connected 3PCF on a  logarithmic 
scale, while $Q(\theta)$ is shown on a linear scale (which can take positive or negative values). 
We can observe first that in general the errors in these measurements are
smaller compared with the ones we showed in Figure 5 for the $u=2$ configurations. As in the $u=2$ configurations,  
there is no significant 
evolution in the amplitude of the correlation values. The 3PCF of equilateral triangles ($u=1$, $\theta\sim 1$)
 is small and even negative, as expected when galaxies cluster in filamentary structures on the largest scales.

\begin{figure*}
\includegraphics[width=12cm]{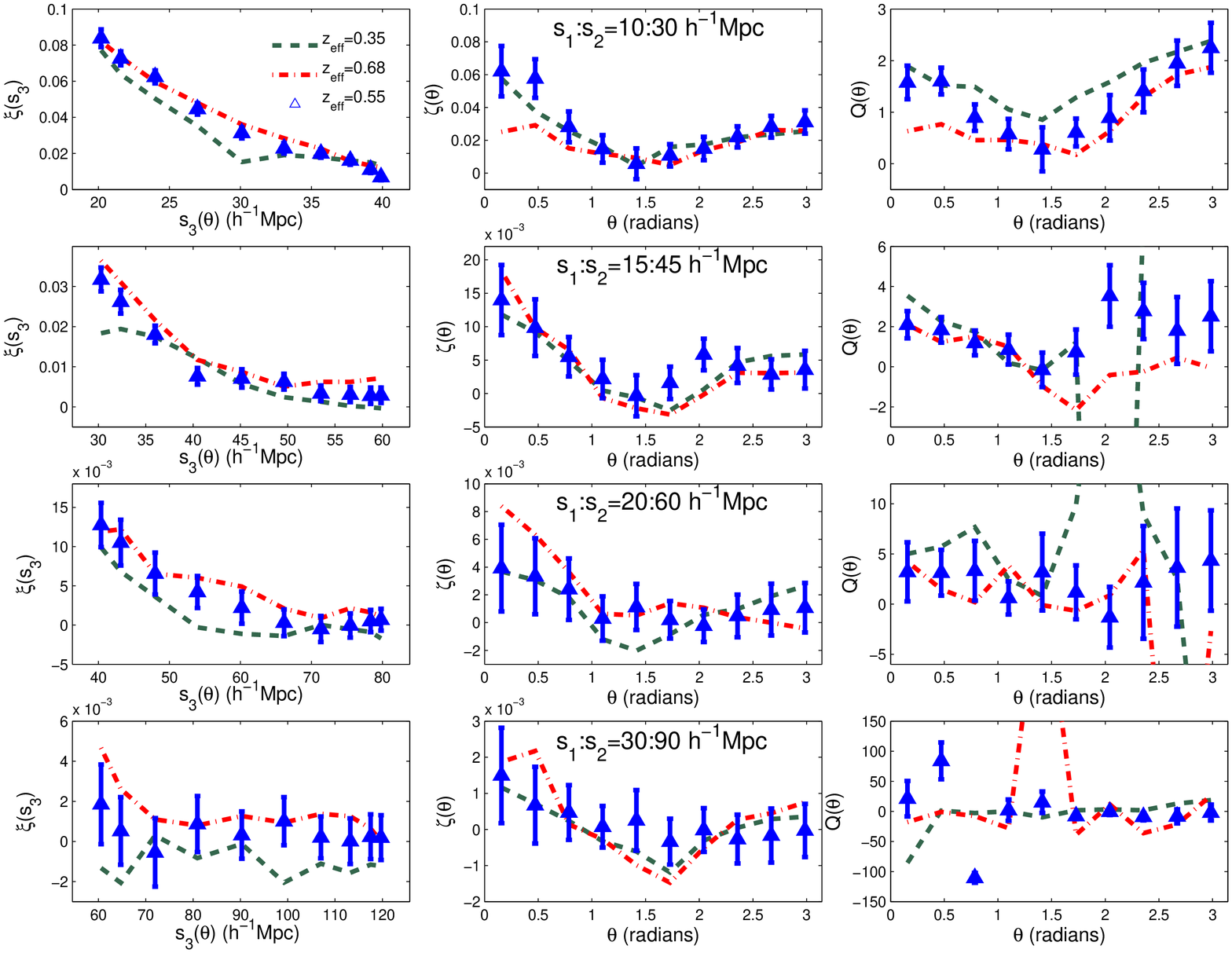}
\caption{\label{fig:wzu3}
The WiggleZ combined correlations of very elongated triangles ($u=3$) 
in redshift-space; for the  2PCF $\xi(s_3(\theta))$ (left), connected 3PCF $\zeta(s,u=3,\theta)$ (middle), and 
reduced 3PCF $Q(s,u=3,\theta)$ (right) of WiggleZ galaxies in the $z_{\rm{eff}}=0.55$ slice (blue triangles),  
in the slice at $z_{\rm{eff}}=0.35$ (green dashed line) and in the $z_{\rm{eff}}=0.68$ slice (red dashed-dotted line).
 Errors have been determined by jack-knife resampling.}
\end{figure*}

In Figure \ref{fig:wzu3} we plot results of the WiggleZ correlation functions measured on very elongated configurations, 
where $u=3$, for the three redshift slices we studied.
For these configurations the signal-to-noise ratio is much smaller than in other
$u$ configurations, specially on large $\theta$. For the   $z_{\rm{eff}}=0.35$ redshift slice (green dashed line), $\zeta$ and $Q$ are
 poorly measured on the largest scales, justifying our decision
to use a maximum separation $s_{3,{\rm max}}$ smaller than that adopted for the other two redshift slices. Notice that we reach scales where the BAO features could in
principle be observed $s_3 \sim 100$ $h^{-1}$Mpc, but the WiggleZ low galaxy bias makes it difficult to achieve a significant detection that could be used to 
constrain cosmological parameters \citep[as claimed by][for SDSS LRGs]{gaztanaga_etal:08a}.

\end{document}